\documentclass[prc,twocolumn,showpacs,floatfix]{revtex4}

\usepackage{amsmath}
\usepackage{graphicx}% Include figure files
\usepackage{bm}% bold math

\begin{document}

%%%%%%%%%%%%%%%%%%
%%% Title Page %%%
%%%%%%%%%%%%%%%%%%
\title{
Coordinate space proton-deuteron scattering calculations   
including Coulomb force effects
}
\author{S.\ Ishikawa}  \email[E-mail:]{ishikawa@hosei.ac.jp}
\affiliation{
Department of Physics, Science Research Center, Hosei University, 
2-17-1 Fujimi, Chiyoda, Tokyo 102-8160, Japan} 

\date{\today}

\begin{abstract}
We present a practical method to solve the proton-deuteron scattering problem at energies above the three-body breakup threshold, in which we treat three-body integral equations in  coordinate space accommodating long-range proton-proton Coulomb interactions. 
The method is examined for phase shift parameters, and then applied to calculations of  differential cross sections in elastic and breakup reactions, analyzing powers, etc. with a realistic nucleon-nucleon force and three-nucleon forces.
Effects of the Coulomb force and the three-nucleon forces on these observables are discussed in comparing with experimental data. 
\end{abstract}

%PACS
%
%21.30.-x 	Nuclear forces (see also 13.75.Cs Nucleon?nucleon interactions)
%21.30.Cb 	Nuclear forces in vacuum
%21.30.Fe 	Forces in hadronic systems and effective interactions

%21.45.-v 	Few-body systems
%21.45.Bc 	Two-nucleon system
%21.45.Ff 	Three-nucleon forces

%23.50.+z 	Decay by proton emission
%25.10.+s 	Nuclear reactions involving few-nucleon systems
%25.40.Cm 	Elastic proton scattering  
%24.70.+s 	Polarization phenomena in reactions

%27.10.+h 	A(less-than-or-equal-to)5  
%13.75.Cs 	Nucleon-nucleon interactions (including antinucleons, deuterons, etc.)
%


\maketitle

%%%%%%%%%%%%%%
%%% Sec. 1 %%%
%%%%%%%%%%%%%%
\section{Introduction}

Scattering observables of three-nucleon ($3N$) systems, proton-deuteron ($pd$) scattering and neutron-deuteron ($nd$) scattering, are good sources of information about unknown aspects of the nuclear forces such as off-shell differences in nucleon-nucleon force (2NF) models, possible evidence of $3N$ forces (3NFs), etc. 
Because of technical advantage of treating charged particles as beam, target, or detected particles over doing neutral ones, available data sets of the $pd$ reaction are richer both in quantity and quality than those of the $nd$ reaction.
On the other hand, due to a mathematical difficulty in treating three-body systems with  long-range Coulomb interactions, a precise calculation of the $pd$ scattering, especially for energies above the three-body breakup threshold (TBT), is one of the most challenging subjects in physics of few-body systems. 

In the last decade, some developments have been made in this problem by calculations based on the Kohn variational principle \cite{Ki99,Ki01} and on the momentum space Faddeev equations \cite{Fa61} using the screening and renormalization method \cite{Al02,De05a,De05b,De06}. 

In this paper, we will present another approach to the $pd$ scattering problem, which is based on integral equations for wave functions in coordinate space. 
Calculations by this approach for non-Coulombnic $3N$ systems with realistic 2NFs and 3NFs 
were performed for ${}^{3}$H in Refs. \cite{Sa86,Is86},  
and for the $nd$ scattering at energies above the TBT in Refs. \cite{Is07,Is07b,Is07c}. 
A direct application of the Faddeev equation to Coulombnic $3N$ systems, namely the ${}^{3}$He bound state and the $pd$ scattering, is known to bring a severe singularity to the integral kernel due to the long-range character of the proton-proton ($pp$) Coulomb force. 
In Ref. \cite{Sa79}, Sasakawa and Sawada proposed a modification of the Faddeev equation to treat the singularity by introducing auxiliary potentials that act between charged spectator particles and the center of mass (c.m.) of the rest pair particles.  
Since we use  an iterative method to solve the integral equations, in which one needs to operate the integral kernel on known functions repeatedly, it is essential to establish an accurate  kernel operation for precise calculations. 
In the integral kernel of this Coulomb-modified Faddeev equation, which will be referred to as the SSF equation, the singularity due to the $pp$ Coulomb potential is expected to be  canceled by the auxiliary potentials, on conditions that three particles are bound or no  three-body breakup channel is open.  
Solutions of the SSF equation were successfully obtained for the ${}^{3}$He bound state in Refs. \cite{Sa81,Wu90,Wu93} and for the $pd$ scattering at energies {\em below} the TBT in Refs. \cite{Is03,Is03b}. 
However, the cancellation is not trivially expected when a three-body breakup channel opens.  
In this paper, we will treat this problem, and show how our approach practically is applicable for $pd$ scattering {\em above} the TBT. 

The next section is devoted to present notations used in this paper and to introduce the SSF equation for the $pd$ scattering in integral equation form.
In Sec.\ \ref{sec:Kernels}, we analyze a problem in the integral kernel of the SSF equation due to the $pp$ Coulomb force, and propose a method to perform numerical calculations. 
Then, some numerical results for $pd$ and $nd$ scattering using a realistic 2NF and 3NFs will be presented in Sec.\ \ref{sec:results}.  
Summary will be given in Sec.\ \ref{sec:summary}.
Our iterative method \cite{Sa86,Is87} is reviewed in Appendix \ref{sec:MCF}, and some useful functions and formulae are appeared in Appendices \ref{sec:functions} and \ref{sec:Green-x}.

%%%%%%%%%%%%%%
%%% Sec. 2 %%%
%%%%%%%%%%%%%%
%%%%%%%%%%%%%%%%%%%%%%%%%%%%%%%%%%%%%%%%%%%%%%%%%%%%%%%%%%%%%%%%%%%%%%%%%%%%%%%%%%%%%%%%%%%
\section{\label{sec:formalism} Three-Body Scattering Equation with Coulomb Force Effects}
%%%%%%%%%%%%%%%%%%%%%%%%%%%%%%%%%%%%%%%%%%%%%%%%%%%%%%%%%%%%%%%%%%%%%%%%%%%%%%%%%%%%%%%%%%

\begin{figure}[tb]
\includegraphics[width=0.4\columnwidth,angle=0]{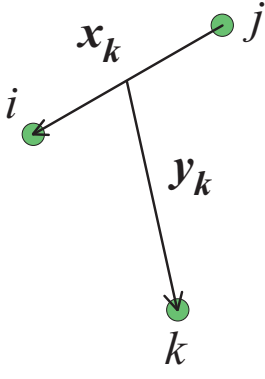}
\caption{(Color online) The Jacobi coordinates of the three-body system.}
\label{fig:Jacobi}
\end{figure}

In this section, we will describe our notations and present the SSF equation by taking a  proton(1)-proton(2)-neutron(3) system as an example.
We will not consider spin's degrees of freedom, angular momentum dependence of the potentials, and 3NFs in describing our formalism because of simplicity.
The deuteron thus is supposed to be a s-wave proton-neutron ($pn$) bound state with energy $E_d$.
We use sets of coordinate systems $\{\bm{x}_k, \bm{y}_k\}$ (the Jacobi coordinates) to describe the three-body system defined as 
\begin{equation}
\bm{x}_k  =  \bm{r}_i - \bm{r}_j, \qquad
\bm{y}_k  =  \bm{r}_k 
   - \frac12 \left(\bm{r}_i + \bm{r}_j \right),
\label{eq:Jacobi}
\end{equation}
where $(i,j,k)$ denote $(1,2,3)$ or their cyclic permutations and $\bm{r}_i$ is the position vector of the particle $i$ (see Fig.\ \ref{fig:Jacobi}). 
Subscripts to indicate particles will be omitted when there is no confusion.
 
We write a three-body Hamiltonian in the c.m. frame as
\begin{equation}
H = H_0 + V_1 + V_2 +  V_3, 
\label{eq:hamiltonian}
\end{equation}
where $H_0$ is the internal kinetic energy operator of the three-body system, 
\begin{equation}
H_0 = T_x(\bm{x})+ T_y(\bm{y}) 
 = -\frac{\hbar^2}{m} \nabla^2_{x} 
 -\frac{3\hbar^2}{4m} \nabla^2_{y}, 
\end{equation}
with nucleon mass $m$, and $V_k$ is a potential to describe the interaction between particles $i$ and $j$ consisting of a short-range nucleon-nucleon potential (2NP) $V_{k}^{S}({x}_{k})$ and  the $pp$ Coulomb potential $V^C(x_{3})$ $ (=e^2/x_3^2)$: 
\begin{equation}
V_{k} = V_{k}^{S}({x}_{k}) + \delta_{k,3} V^C(x_{3}).
\end{equation}

%%%%%%%%%%%%%%%%%%%%%%%%%%%%%%%%%%%
%%%%%%%%%%%%%%%%%%%%%%%%%%%%%%%%%%%

We begin with a differential form of the SSF equation for a three-body c.m. energy $E (>0)$ \cite{Sa79},
\begin{subequations}
\label{eq:mod-fad-dif}
\begin{eqnarray}
\left[ E - H_0 - V^{S}_{1} - u^{C}(y_{1}) \right] \Phi_{1}
 &=& \left[ \Delta \Phi \right]_1,
\label{eq:mod-fad-dif-1}
\\
\left[ E - H_0 - V^{S}_{2} - u^{C}(y_{2}) \right] \Phi_{2}
 &=& \left[ \Delta \Phi \right]_2,
\label{eq:mod-fad-dif-2}
\\
\left[ E - H_0 - V^{S}_{3} - V^C(x_{3}) \right] \Phi_{3}
 &=& \left[ \Delta \Phi \right]_3,
\label{eq:mod-fad-dif-3}
\end{eqnarray}
\end{subequations}
where $\Phi_k$'s are Faddeev components, and $u^{C}(y_{k})$ is an auxiliary Coulomb potential acting between the particle $k$ and the c.m. of the pair $i j$,
\begin{equation}
 u^{C}(y_k) = \frac{e^2}{y_k} \quad (k=1, 2). 
\end{equation}
The symbols $\left[ \Delta \Phi\right]_k$ in the right hand side denote  
\begin{equation}
 \left[ \Delta \Phi \right]_k \equiv 
\left\{
\begin{array}{l}
V_k^S \left( \Phi_i + \Phi_j \right)
\qquad \qquad  (k=1,2),
\\
~
\\
V_3^S \left( \Phi_1 + \Phi_2 \right) + \left( V^C(x_{3})-u^C(y_1) \right) \Phi_1
\\
 + \left( V^C(x_{3})-u^C(y_2) \right) \Phi_2 
\qquad (k=3).
\end{array} \right.
\end{equation}

The auxiliary potentials play different roles on each side of Eqs. (\ref{eq:mod-fad-dif-1})-(\ref{eq:mod-fad-dif-3}).
On the left hand side of Eqs. (\ref{eq:mod-fad-dif-1}) and (\ref{eq:mod-fad-dif-2}), these potentials work to distort the spectator proton from a free state (see Eq. (\ref{eq:spectator_F}) below). 
On the right hand side of Eq. (\ref{eq:mod-fad-dif-3}), i.e. in $\left[ \Delta \Phi \right]_3$, the auxiliary potential is expected to cancel out the long-rangeness of the $pp$ Coulomb potential $V^C$.
The latter point will be discussed later in this article. 
It should be noted that the auxiliary potentials in Eqs. (\ref{eq:mod-fad-dif-1})-(\ref{eq:mod-fad-dif-3}) are eliminated when all the equations are summed up, which makes the sum $\Phi_1+\Phi_2+\Phi_3$ an eigenstate of the Hamiltonian  (\ref{eq:hamiltonian}).

We will consider a $pd$ scattering state of the initial $pd$ momentum $\bm{p}_0$, which gives the three-body c.m. energy
\begin{equation}
E =  E_{p_0} + E_d,
\end{equation} 
where
\begin{equation}
E_{p_0} = \frac{3\hbar^2}{4m} p_0^2.
\label{eq:E_p0}
\end{equation}
Integral form of the SSF equation, whose formal derivation is given in Ref. \cite{Is03}, is
\begin{equation}
\Phi_k = \bar{\delta}_{k,3}  
 \Phi^d(\bm{x}_k) F^C(\bm{y}_k;\bm{p}_0,\eta_0)  +{\cal G}_k(E)  \left[ \Delta \Phi \right]_k, 
\label{eq:mod-fad-int}
\end{equation}
where $\bar{\delta}_{i,j} = 1-\delta_{i,j}$; 
$\Phi^d(\bm{x})$ is the deuteron state;
$F^C(\bm{y},\bm{p}_0,\eta_0)$ is a scattering state in the Coulomb potential $u^C(y)$, which satisfies 
\begin{equation}
\left[ T_y(\bm{y}) + u^C(y) \right] F^C(\bm{y};\bm{p}_0,\eta_0) 
   = E_{p_0} F^C(\bm{y};\bm{p}_0,\eta_0) 
\end{equation}
with the Coulomb parameter $\eta_0 = \eta(p_0)$ given by Eq.\ (\ref{eq:eta_p});
the operators ${\cal G}_k(E)$ are channel Green's functions defined as
\begin{equation}
{\cal G}_k (E) \equiv 
\left\{
 \begin{array}{l}
 \displaystyle{\frac1{ E+\imath\varepsilon - H_0 - V^S_k - u^C(y_k)} }
\\
  \qquad \qquad \qquad \qquad\qquad (k=1, 2), 
\\
 \displaystyle{\frac1{ E+\imath\varepsilon - H_0 - V^S_3-V^C(x_3)} }
\\
  \qquad \qquad \qquad \qquad\qquad (k=3),
\end{array}
\right.
\end{equation}
with the parameter $\varepsilon$ being a small positive number to give outgoing waves.

%%%%%%%%%%%%%%%%%%%%%%%%%%%%%%%%%%%%%%%%%%%
%%% Partial wave and spectator function %%%
%%%%%%%%%%%%%%%%%%%%%%%%%%%%%%%%%%%%%%%%%%%
A partial-wave decomposition is performed by introducing an angular function denoted as
$\vert \alpha (\hat{\bm{x}},\hat{\bm{y}}) )$, 
\begin{equation}
%{\cal Y}_\alpha (\hat{\bm{x}},\hat{\bm{y}})
\left\vert \alpha (\hat{\bm{x}},\hat{\bm{y}}) \right)
 = \left[ Y_L(\hat{\bm{x}}) \otimes Y_\ell(\hat{\bm{y}}) \right]_{M_0}^{J_0},
\end{equation}
where $\bm{L}$ denotes the relative orbital angular momentum of the pair particles; 
$\bm{\ell}$ the orbital angular momentum of the spectator particle; 
$\bm{J}_0 (= \bm{L}+\bm{\ell})$ and $M_0$ the total angular momentum of the three particles and its third component, respectively.
The set of the quantum numbers ($L, \ell, J_0, M_0$) is represented by the index $\alpha$.

We define complete orthogonal sets of functions describing the angular parts of the three-body system with a state index $\alpha$ and the radial part of the spectator particle with momentum $p$ and angular momentum $\ell$, 
\begin{equation}
\left\vert {\cal F}_{k,\underline{\alpha}} \right)
 \equiv \left\vert \alpha (\hat{\bm{x}}_k,\hat{\bm{y}}_k) \right) 
 \times 
\left\{ \begin{array}{l} 
\sqrt{\frac2{\pi}} \frac{F_\ell(\eta(p), p y_k)}{y_k} 
\qquad(k =1, 2), 
\cr
\\ 
\sqrt{\frac2{\pi}} p j_\ell (p y_k)
\qquad 
\qquad(k =3), 
\end{array} \right.
\label{eq:spectator_F}
\end{equation}
where $F_\ell(\eta,r)$ is the regular Coulomb function of Eq. (\ref{eq:Coulomb-partial}) \cite{Me65,Ab65}, $\eta(p)$ is the Coulomb parameter of Eq.\ (\ref{eq:eta_p}),  
and  $j_\ell(r)$ is the spherical Bessel function.
The underline implies a dependence on the momentum $p$.
These functions satisfy a complete relation,
\begin{equation}
\int_{\underline{\alpha}}~
 \left\vert {\cal F}_{k,\underline{\alpha}} \right)
  \left({\cal F}_{k,\underline{\alpha}} \right\vert 
= 1, 
\end{equation}
and an orthogonal relation,
\begin{equation}
\left( {\cal F}_{k,\underline{\alpha}} 
 \vert {\cal F}_{k,\underline{\alpha}^\prime}\right) = 
   \delta_{\alpha,\alpha^\prime} \delta(p-p^\prime)=\delta_{\underline{\alpha},\underline{\alpha}^\prime} , 
\end{equation}
where $\int_{\underline{\alpha}}$ means $\displaystyle{\sum_\alpha} \int_0^\infty dp$, 
and $(~|~)$ integrations over the variables, $\hat{\bm{x}}$ and $\bm{y}$.

The channel Green's function is decomposed by the complete set to give,  
\begin{equation}
{\cal G}_k(E)  = \int_{\underline{\alpha}} 
 \left\vert {\cal F}_{k,\underline{\alpha}} \right) G_{k,\underline{L}}
   \left( {\cal F}_{k,\underline{\alpha}} \right\vert.
\label{eq:ch-Gfn_3}
\end{equation}
Here, $G_{k,\underline{L}}$ is a two-body Green's operator, 
\begin{equation}
G_{k,\underline{L}} = 
\frac1{E_q + \imath \varepsilon - T_{L}(x) - V_k^S(x) - \delta_{k,3} V^C(x)}, 
\label{eq:Green_3}
\end{equation}
where $E_q$ is the energy of the two-body subsystem given by 
\begin{equation}
E_q = E - \frac{3 \hbar^2}{4m} p^2 = \frac{\hbar^2}{m} q^2,
\label{eq:Eq}
\end{equation}
and 
\begin{equation}
T_L(x) = -\frac{\hbar^2}{m} \left(
   \frac{d^2}{dx^2} + \frac2x \frac{d}{dx} - \frac{L(L+1)}{x^2} \right).
\end{equation}

%%%%%%%%%%%%%%
%%% Sec. 3 %%%
%%%%%%%%%%%%%%
%%%%%%%%%%%%%%%%%%%%%%%%%%%%%%%%%%%%%%%%%%%%%%%%%%%%%%%%%%%%%%%%%%%%%%%%%%%%%%
\section{\label{sec:Kernels} Coulomb force effects in the SSF integral kernel}
%%%%%%%%%%%%%%%%%%%%%%%%%%%%%%%%%%%%%%%%%%%%%%%%%%%%%%%%%%%%%%%%%%%%%%%%%%%%%%
 
The SSF integral equation presented in the previous section has a form of the inhomogeneous linear equation.
We are going to solve this by applying an iterative method developed in Refs. \cite{Sa86,Is87} (and references therein), which is called as the Method of Continued Fractions (MCF).
In general, iterative methods to solve an linear equation require to operate the kernel to functions that are given in preceding iterative steps. 
The MCF algorithm, which is reviewed in Appendix \ref{sec:MCF}, also includes such operations as indicated in Eqs. (\ref{eq:MCF-Fi}) and (\ref{eq:MCF-Gi}).
Calculations of the SSF integral kernel consist of two parts: 
a particle exchange operation and the operation of the Green's functions.
Some technical notes of the former part are given in Refs. \cite{Sa82,Sa83}, 
and those of the latter part for the $nd$ scattering above the TBT  in Ref. \cite{Is07}, 
which are useful also in the $pd$ scattering. 
In this section, we will study some problems of the SSF integral kernel proper to the $pd$ scattering problem.

%%%%%%%%%%%%%%%%%%%%%%%%%%%%%%%%
\subsection{SSF integral kernel}
%%%%%%%%%%%%%%%%%%%%%%%%%%%%%%%%
Let us consider to operate the SSF kernel on given functions $\Phi_k(\bm{x},\bm{y})$: 
\begin{equation}
\Theta_k(\bm{x}_k,\bm{y}_k) \equiv  {\cal G}_k(E) \left[ \Delta \Phi \right]_k.
\label{eq:Theta-def}
\end{equation}

The channel Green's function for $k=1$ or $2$, where the pair is a $pn$ system, possesses a pole corresponding to the deuteron bound state. 
In order to treat this pole, we apply a standard subtraction method, 
in which we use an identity, 
\begin{equation} 
1 = \sum_{\alpha_0} \left\vert \alpha_0 \phi^d\right> \left< \phi^d\alpha_0 \right\vert
   +      \left[ 1 -  \sum_{\alpha_0} \left\vert \alpha_0 \phi^d\right>
            \left< \phi^d \alpha_0 \right\vert  \right],
\end{equation} 
where $\phi^d(x)$ is the radial part of the deuteron wave function with orbital angular momentum $L_0 (=0)$, and the index $\alpha_0 = (L_0,\ell_0,J_0,M_0)$  denotes the three-body partial wave states that couple to the two-body state with $L_0$.
By applying the identity to ${\cal G}_k$, we obtain
\begin{eqnarray}
{\cal G}_k(E) &=&  
\sum_{\alpha_0} \left\vert \alpha_0 \phi^d \right> \breve{G}_{C,\ell_0}(E_{p_0})
    \left< \phi^d \alpha_0 \right\vert
\cr
& &+ \int_{\underline{\alpha}} \left\vert {\cal F}_{k,\underline{\alpha}} \right)
  G_{k,\underline{L}} 
\left( {\cal F}_{k,\underline{\alpha}}  \right\vert
\cr
& &- \int_{\underline{\alpha}_0}  
  \left\vert {\cal F}_{k,\underline{\alpha}_0} \phi^d \right>
      \frac1{E_q - E_d}  \left< \phi^d {\cal F}_{k,\underline{\alpha}_0} \right\vert 
\cr
&& \qquad\qquad\qquad\qquad (k=1,2).
\label{eq:ch-Gfn_12}
\end{eqnarray} 
Here, $\breve{G}_{C,\ell_0}(E_{p_0})$ is the partial wave component of the Coulomb Green's function for the outgoing proton, 
\begin{equation}
\breve{G}_{C,\ell_0}(E_{p_0}) \equiv
  \frac1{E_{p_0} + \imath \varepsilon - T_{\ell_0}(y) - u^C(y)}
\end{equation}
with
\begin{equation}
T_{\ell}(y) = -\frac{3\hbar^2}{4m} \left(
   \frac{d^2}{dy^2} + \frac2y \frac{d}{dy} - \frac{\ell(\ell+1)}{y^2} \right).
\end{equation}

The function $\Theta_k(\bm{x},\bm{y})$ thereby can be written as
\begin{eqnarray}
&& \Theta_k(\bm{x},\bm{y})
\cr
&&
\cr
&=& \left\{ \begin{array}{l}
\displaystyle{\sum_{\alpha_0}} \left\vert \alpha_0 \right) \phi^d(x) {\eta}^{(e)}_{k,\alpha_0}(y)
\\
 + \displaystyle{ \int_{\underline{\alpha}} }\left\vert {\cal F}_{k,\underline{\alpha}} \right)
  \left\{ \theta_{k,\underline{\alpha}}(x) 
   - \delta_{\alpha,\alpha_0}\phi^d(x) C_{k,\underline{\alpha}_0} \right\}
\\
 \qquad\qquad\qquad \qquad\qquad(k=1,2),
\\
\displaystyle{\int_{\underline{\alpha}} }
 \left\vert {\cal F}_{k,\underline{\alpha}} \right) 
 \theta_{k,\underline{\alpha}}(x)
  \qquad\qquad(k=3).
\end{array} \right.
\label{eq:theta-xy}
\end{eqnarray}

Here, the function ${\eta}^{(e)}_{k,\alpha_0}(y)$ $(k=1,2)$ represents an elastic component in the scattering, 
\begin{equation}
{\eta}^{(e)}_{k,\alpha_0}(y)
 =  \int_0^\infty y^{\prime 2} dy^\prime 
  \breve{G}_{C,\ell_0}(y, y^\prime;E_{p_0}) \omega^{(e)}_{k,\alpha_0}(y^\prime)
\label{eq:phi-e}
\end{equation}
with 
\begin{equation}
\breve{G}_{C,\ell}(y, y^\prime;E_{p}) \equiv \langle y \vert \breve{G}_{C,\ell}(E_{p})\vert y^\prime \rangle,
\label{eq:breve-G}
\end{equation}
 and a source function $\omega^{(e)}_{k,\alpha_0}(y)$ given by
\begin{eqnarray}
 \omega^{(e)}_{k,\alpha_0}(y)
&=&  \left< \phi^d \alpha_{0} \vert \left[ \Delta \Phi \right]_k \right> 
\cr
&=&  \left< \phi^d \alpha_{0} \left\vert V_k^S \right\vert \Phi_i + \Phi_j  \right>.
\label{eq:omega_y}
\end{eqnarray}

The explicit expression of the Green's function $\breve{G}_{C,\ell}(y, y^\prime;E_{p})$,  Eq. (\ref{eq:brG-l}) \cite{Me65}, gives the asymptotic form of ${\eta}^{(e)}_{k,\alpha_0}(y) $  as
\begin{equation}
{\eta}^{(e)}_{k,\alpha_0}(y) \mathop{\to}_{y \to \infty} 
  \frac{e^{\imath \sigma_{\ell_0}(\eta_0)} u_{\ell_0}^{(+)}(\eta_0,p_0 y)}{y} 
 T^{(e)}_{k,\alpha_0},
\label{eq:phi_e_asym}
\end{equation}
where $T^{(e)}_{k,\alpha_0}$ is an amplitude defined by
\begin{equation}
T^{(e)}_{k,\alpha_0} = - \left( \frac{4m}{3 \hbar^2  p_0}\right) 
    \int_0^\infty dy  {F_\ell(\eta_0,p_0y)}
  y \omega^{(e)}_{k,\alpha_0}(y).
\end{equation}
Above the TBT, the source function $\omega^{(e)}_{k,\alpha_0}(y)$ reveals a long-range behavior of ${\cal O}(y^{-5/2})$ even in the case of the $nd$ scattering due to the particle exchange with breakup channel.
This property was studied to develop a numerical treatment in Ref. \cite{Is07}.

The coefficient $C_{k,\underline{\alpha}_0}$ $(k=1,2)$ and the function $\theta_{k,\underline{\alpha}}(x)$ $(k=1,2,3)$ in Eq. (\ref{eq:theta-xy}) are defined as  follows:
\begin{equation}
C_{k,\underline{\alpha}_0} = \frac1{E_q - E_d} 
   \left\langle \phi^d  {\cal F}_{k,\underline{\alpha}_0} \vert  V_k^S \vert \Phi_i + \Phi_j  \right\rangle,
\end{equation}
\begin{equation}
\theta_{k,\underline{\alpha}}(x)
 = \langle x \vert G_{k,\underline{L}} \vert \omega_{k,\underline{\alpha}} \rangle,
\label{eq:theta-Gomega}
\end{equation}
where a source function $\omega_{k,\underline{\alpha}}(x)$ is composed of a contribution from the short-range potential and one from the Coulomb potentials, 
\begin{equation}
\omega_{k,\underline{\alpha}}(x_k) 
= \omega_{k,\underline{\alpha}}^S(x_k) + \delta_{k,3} \omega_{\underline{\alpha}}^C(x_k) 
\label{eq:omega_x}
\end{equation}
with
\begin{eqnarray}
\omega_{k,\underline{\alpha}}^S(x_k) 
&=& \left( {\cal F}_{k,\underline{\alpha}} \left\vert V_k^S \right\vert \Phi_i + \Phi_j \right\rangle
\cr
&=& V_k^S(x_k) \left( {\cal F}_{k,\underline{\alpha}} \vert \Phi_i + \Phi_j \right\rangle
\label{eq:omega-S},
\\
\omega_{\underline{\alpha}}^C(x_3) 
&=&  \left( {\cal F}_{3,\underline{\alpha}} \left\vert
   V^C(x_3)-u^C(y_1) \right\vert \Phi_1 \right\rangle
\cr
&+&  \left( {\cal F}_{3,\underline{\alpha}} \left\vert
   V^C(x_3)-u^C(y_2) \right\vert \Phi_2 \right\rangle.
\label{eq:omega-C}
\end{eqnarray}

Note that the apparent singularity of $C_{k,\underline{\alpha}_0}$ for $E_q=E_d$, or  $p=\sqrt{\frac{4m}{3\hbar^2} \left(E+\vert E_d\vert\right)}$, (see Eq. (\ref{eq:Eq})) is canceled by that of the function $\theta_{k,\underline{\alpha}}(x)$ arising from the two-body Green's function, $G_{k,\underline{L}}$, and therefore the standard quadrature can be applied to perform the $p$-integration of Eq.\ (\ref{eq:theta-xy}) as far as both terms are treated together as demonstrated in Ref. \cite{Is07}.

In actual calculations of the functions $\theta_{k,\underline{\alpha}}(x)$, we consider an ordinary differential equation that is transformed from Eq.\ (\ref{eq:theta-Gomega}),
\begin{equation}
\left[ E_q  - T_{L}(x) - V_k^S(x) - \delta_{k,3} V^C(x)\right] \theta_{k,\underline{\alpha}}(x) 
 = {\omega}_{k,\underline{\alpha}}(x).
\label{eq:Sch-Fad}
\end{equation}
A boundary condition to get physical solution of this equation depends on energy of the two-body sub-system $E_q$, and thus on the integral variable $p$ in Eq.\ (\ref{eq:theta-xy}) via Eq. (\ref{eq:Eq}). 
According to the sign of $E_q$, the range of $p$ ($0 \le p < \infty$) is divided into two regions:
(i) $0 \le p \le p_c = \sqrt{\frac{4m}{3\hbar^2}E}$, where $E_q\ge0$, and 
(ii)  $p_c < p < \infty$, where $E_q<0$.
Corresponding boundary conditions for $k=1,2$ are:
\begin{equation}
\theta_{k,\underline{\alpha}}(x) 
   \mathop{\propto}_{x \to \infty}
\left\{
\begin{array}{ll}
  h^{(+)}_{L} ( q x) & (0 \le p \le p_c),
\\
~
\\
  h^{(+)}_{L} ( \imath \vert q \vert x)& ( p_c < p < \infty),
\end{array}
\right.
\label{eq:bd-cond_12}
\end{equation}
where $h^{(+)}_\ell(r)$ is the spherical Hankel function with the outgoing wave.
For $k=3$, where the $pp$ Coulomb potential is acting, we have 
\begin{equation}
\theta_{3,\underline{\alpha}}(x) 
  \mathop{\propto}_{x \to \infty}
\left\{
\begin{array}{ll}
  \frac{u^{(+)}_{L}(\gamma(q),qx)}{x} & (0 \le p \le p_c),
\\
~
\\
  \frac{W_{-\gamma(\vert q\vert),L+1/2} ( 2 \vert q \vert x)}{x}& ( p_c < p < \infty),
\end{array}
\right.
\label{eq:bd-cond_3}
\end{equation}
where
$\gamma(q)$ is given by Eq. (\ref{eq:gamma}) and 
 $W_{\kappa, \mu}(z)$ is the Whittaker function \cite{Ab65}.
We solve Eq. (\ref{eq:Sch-Fad}) with above conditions by applying usual techniques as in the two-body problem, e.g. the Numerov algorithm \cite{Sa83}.
Treatments of Eq. (\ref{eq:Sch-Fad}) for $k=1,2$ in the region (i), which are same as for the $nd$ scattering, are described in the Appendix B of Ref. \cite{Is07}.
While those for the $k=3$ case, where we need to consider Coulomb force effects, are given in  Appendix \ref{sec:Green-x} of this article.

The asymptotic form of the function $\Theta_{k}(\bm{x},\bm{y})$ is obtained by evaluating 
Eq. (\ref{eq:theta-xy}) with the saddle-point approximation \cite{Sa77,Ru65} together with an explicit asymptotic form of $\theta_{3,\underline{\alpha}}(x)$ for $0 \le p \le p_c$  given by Eq. (\ref{eq:theta_asym}). 
We notice that the Coulomb force effects appear in the spectator variable $\bm{y}_k$ for $k=1,2$ and in the pair coordinate $\bm{x}_k$ for $k=3$.
The result is
\begin{eqnarray}
\Theta_{k}(\bm{x},\bm{y}) 
&\displaystyle{\mathop{\to}_{x \to \infty}} &
- e^{\frac{\pi}{4}\imath}  \sum_{\alpha} \vert \alpha )
 \imath^{-L-\ell} \left(\frac{4K_0}{3}\right)^{3/2}
\cr
&& \times \frac{e^{\imath \left( K_0 R - \delta_{k,3}\gamma(\bar{q}) \ln(2\bar{q}x) 
 - \bar{\delta}_{k,3} \eta(\bar{p}) \ln(2\bar{p}y) \right)}}{R^{5/2}}
\cr
&& \times  B_{k,\alpha}(\Theta),
\label{eq:phi_asym}
\end{eqnarray}
where the limit is considered to be taken with $x/y$ being fixed, a hyper radius $R$ and a hyper angle $\Theta$ are introduced as
\begin{equation}
R = \sqrt{x^2 + \frac43 y^2},
\end{equation}
\begin{equation}
x = R \cos\Theta,~~~~y= \sqrt{\frac34} R \sin\Theta,
\end{equation}
$K_0$ and the momenta, $\bar{q}$ and $\bar{p}$, are given by
\begin{equation}
K_0 = \sqrt{\frac{m}{\hbar^2} E}, 
\end{equation}
\begin{equation}
 \bar{q} = K_0 \cos\Theta, \qquad \bar{p}= \sqrt{\frac43} K_0 \sin\Theta.
\end{equation}

Here, $B_{k,\alpha}(\Theta)$ is a breakup amplitude defined as
\begin{equation}
B_{k,\alpha}(\Theta) = 
  - \frac1{\bar{p}} \frac{m}{\hbar^2} \frac{1}{1-\imath {\cal K}_{L}(\bar{q})}
  \langle \bar{\psi}_{k,L}(\bar{q})\vert
   \omega_{k,\underline{\alpha}} \rangle,
\label{def:B_amp}
\end{equation}
where $\bar{\psi}_{k,L}(x;q)$ is a two-body scattering solution with the standing wave boundary condition and ${\cal K}_{L}(q)$ is a scattering $K$-matrix for the two-body scattering (see Appendix \ref{sec:Green-x}). 

%%%%%%%%%%%%%%%%%%%%%%%%%%%%%%%
\subsection {Coulomb long range effects}
%%%%%%%%%%%%%%%%%%%%%%%%%%%%%%%
In solving the differential equation (\ref{eq:Sch-Fad}) numerically, we need to set a value $x_M$ by a condition that the source function $\omega_{k,\underline{\alpha}}(x)$ should vanish so that the solution reaches its asymptotic form given by Eqs. (\ref{eq:bd-cond_12}) or (\ref{eq:bd-cond_3}) for $x>x_M$. 
The range of $\omega_{k,\underline{\alpha}}(x)$ thus is an important issue in our calculations. 
Eq. (\ref{eq:omega-S}) shows that the range of the short-range potential term $\omega_{k,\underline{\alpha}}^S(x)$ for $k=1,2,3$ is determined by the range of $V_k^S(x)$. 
Therefore, in the case of $k=1, 2$, where there is no contribution from the Coulomb term, we set $x_M$ to be a value larger than the range of the 2NP, e.g. 10 fm.

\begin{figure}[tb]
\includegraphics[width=0.5\columnwidth,angle=0]{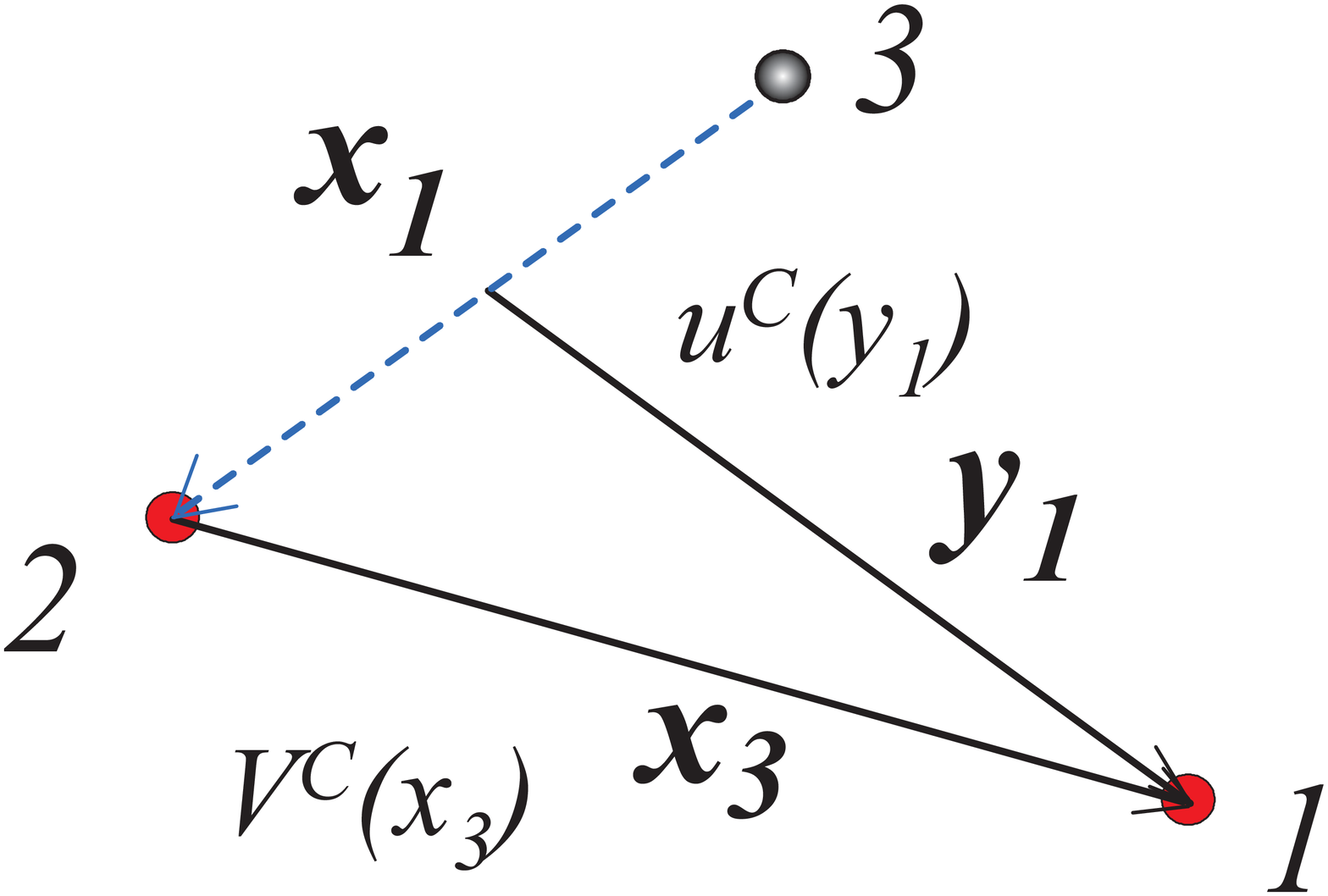}
\caption{(Color online) 
Jacobi coordinates ($\bm{x}_1$, $\bm{y}_1$, and $\bm{x}_3$) to describe the $pp$ Coulomb potential $V^C(x_3)$ and the auxiliary Coulomb potential $u^C(y_1)$. 
} 
\label{fig:auxCoul}
\end{figure}

In the case of $k=3$ on the other hand, the source function includes the Coulomb term  $\omega_{\underline{\alpha}}^C(x_3)$, whose range depends on a factor
\begin{eqnarray*}
&&\left\{ V^C(x_3) - u^C(y_1) \right\} \Phi_1(\bm{x}_1,\bm{y}_1) 
\quad + \quad (1 \leftrightarrow 2).
\\
&=& \left( \frac1{x_3} - \frac1{y_1} \right) \Phi_1(\bm{x}_1,\bm{y}_1) 
\quad + \quad (1 \leftrightarrow 2).
\end{eqnarray*} 
In our iterative scheme (see Appendix \ref{sec:MCF}), the zeroth order of the source function  $\omega_{3,\underline{\alpha}}^{[0]}(x_3)$ is calculated by putting    $\Phi_1(\bm{x}_1,\bm{y}_1)=\Phi^d(\bm{x}_1) F^C(\bm{y}_1;\bm{p}_0,\eta_0)$, in which the magnitude of the variable $\bm{x}_1$ is restricted within the range of the deuteron size. 
Using an expression given by the definition of the Jacobi coordinates, Eq.\ (\ref{eq:Jacobi}),  (see also  Fig. \ref{fig:auxCoul}),  
\begin{equation}
\bm{y}_1 =  \bm{x}_3 + \frac12 \bm{x}_1,  
\end{equation}
we can easily show that 
\begin{equation}
\frac1{x_3} - \frac1{y_1} 
  = \frac1{x_3} - \frac1{ \vert \bm{x}_3 + \frac12 \bm{x}_1 \vert } 
    \mathop{\to}_{x_3 \to \infty} {\cal O}(x_3^{-2}). 
\label{eq:x3-y1}
\end{equation} 
The same situation holds for the replacement of $(1 \leftrightarrow 2)$. 
The source term $\omega_{\underline{\alpha}}^C(x_3)$ therefore supposed to be a short-range function due to a cancellation between $V^C$ and $u^C$.

\begin{figure}[tb]
\includegraphics[width=0.45\columnwidth,angle=0]{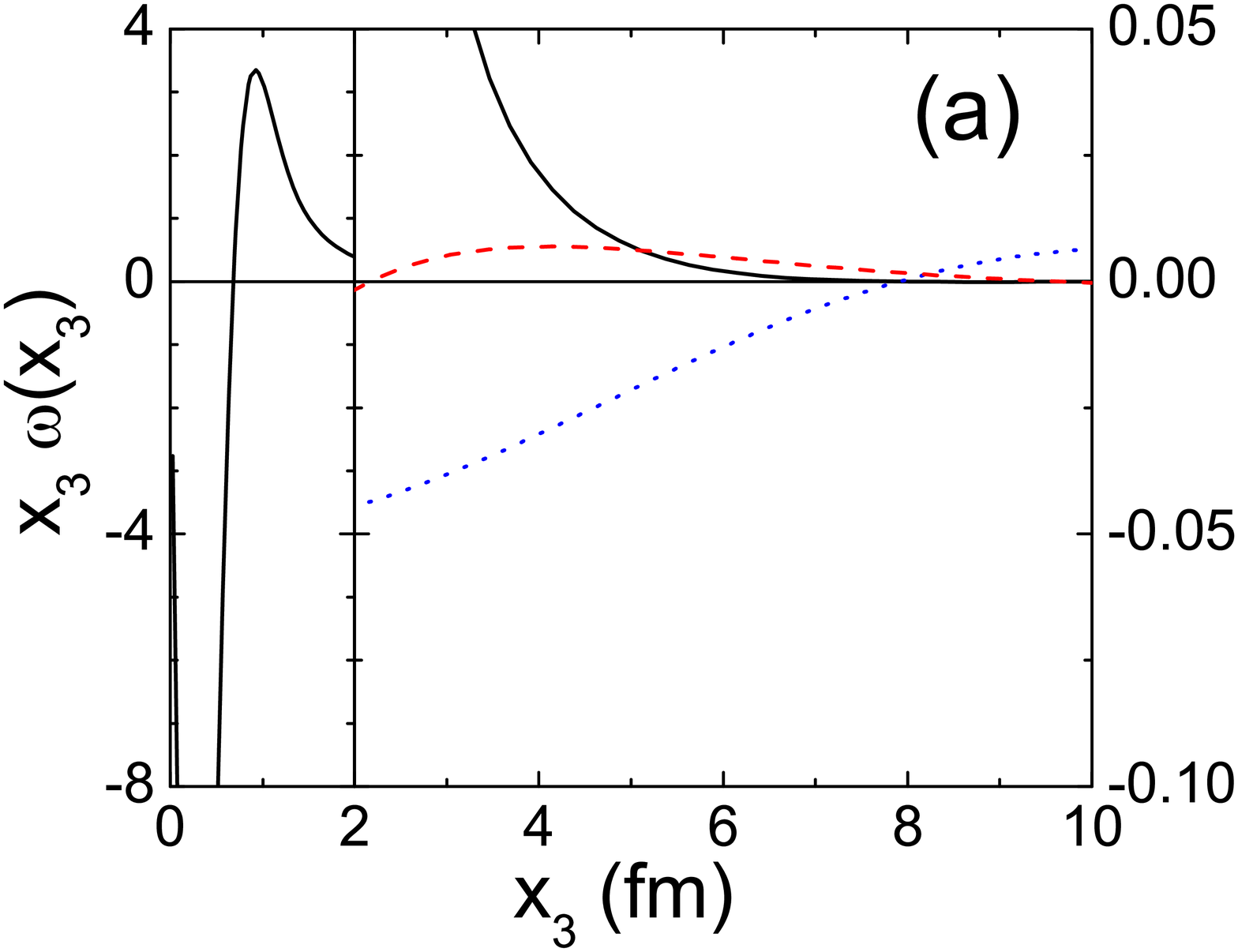}
\includegraphics[width=0.45\columnwidth,angle=0]{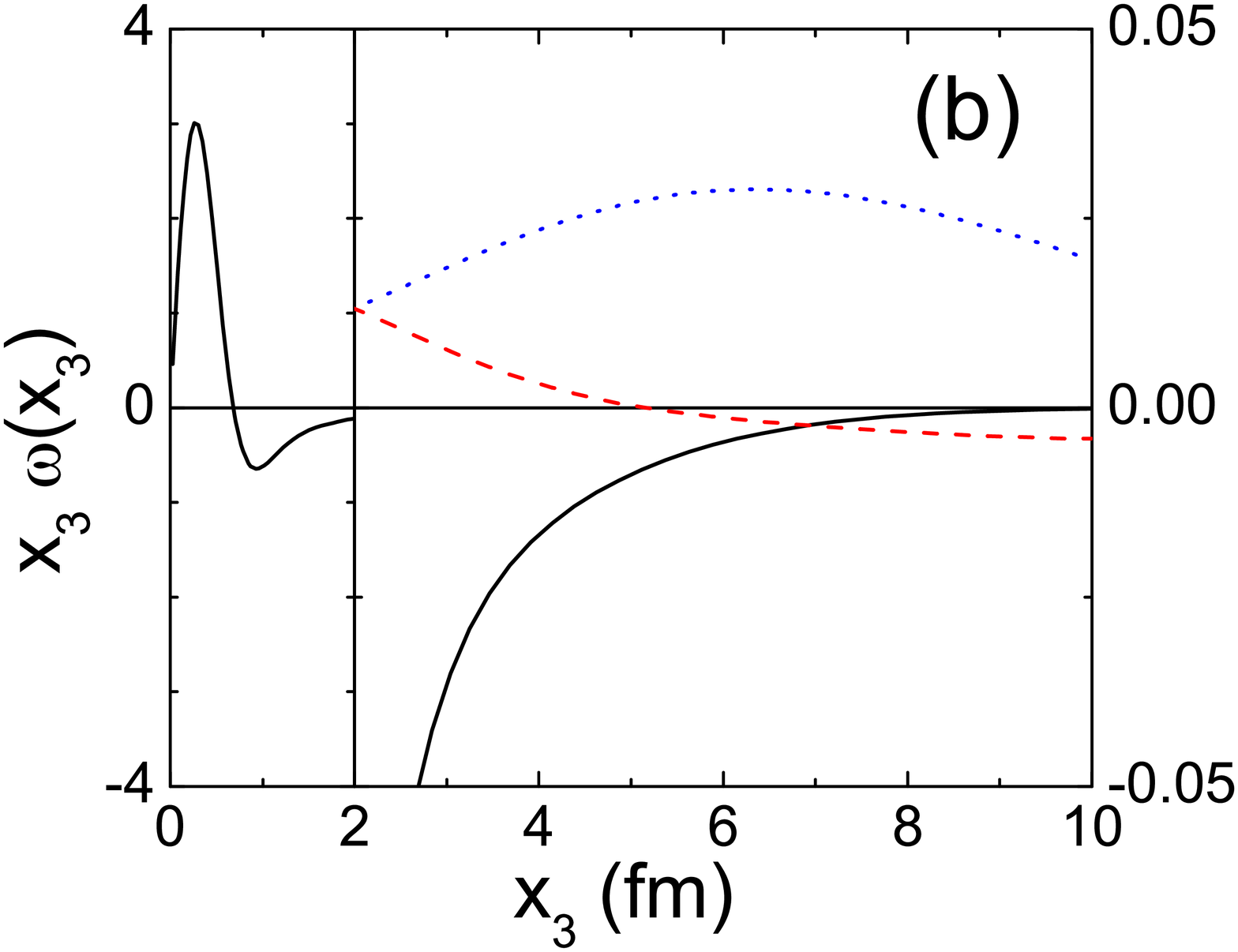}
\caption{(Color online) 
Examples of the source functions (a) for the zeroth order and (b) for the first order calculations. 
The solid curves denote the contribution from the short-range 2NP  $\omega_{3,\underline{\alpha}}^S(x_3)$, the dotted curves that from the Coulomb potentials  $\omega_{\underline{\alpha}}^C(x_3)$ neglecting the contribution from $u^C$, and the dashed curves $\omega_{\underline{\alpha}}^C(x_3)$ including the contribution from both of $V^C$ and $u^C$. 
Note that the scales of the vertical axis change at $x_3=$ 2.0 fm.
} 
\label{fig:omega-xp}
\end{figure}

An example of the cancellation is shown in Fig. \ref{fig:omega-xp} (a),
where we plot components of $\omega_{3,\underline{\alpha}}^{[0]}(x_3)$ for a partial wave state of ${}^{1}$S$_{0}(pp)-s_{1/2}$ and the total three-body angular momentum and parity of $1/2^+$, and $p=0.30$ fm$^{-1}$ for the $pd$ scattering at incident proton energy $E_p$ = 13.0 MeV using the Argonne V$_{18}$ (AV18) 2NP  \cite{Wi95}.
In the figure, a component due to the 2NP, $\omega_{3,\underline{\alpha}}^{S}(x_3)$ (the solid  curve), and components due to the Coulomb potentials, 
$\omega_{\underline{\alpha}}^C(x_3)$, including only $V^C$ (the dotted curve) and both of $V^C$ and $u^C$ (the dashed curve) are plotted.
For $x_3<$ 2 fm, only the term $\omega_{3,\underline{\alpha}}^{S}(x_3)$ is plotted since the Coulomb contributions are very small in this region.  
As shown by the dotted and dashed curves in the figure, the contribution of $V^C$ is well canceled by that of $u^C$ for large values of $x_3$  in the zeroth order calculation.

%%%%%%%%%%%%%%%%%%%%%%%%%%%%%%%%%%%%%%%%%%%%%%%%%%%%%%%%%%%%%%%%%%%%%%%%%%%%%%%
%%%%%%%%%%%%%%%%%%%%%%%%%%%%%%%%%%%%%%%%%%%%%%%%%%%%%%%%%%%%%%%%%%%%%%%%%%%%%%%

On the other hand, Fig. \ref{fig:omega-xp} (b) shows the components of $\omega_{3,\underline{\alpha}}^{[1]}(x_3)$  
calculated from functions $\Phi_k^{[1]}(\bm{x}_k,\bm{y}_k)$ that are obtained by the operation 
of the kernel to the initial state. 
Once the integral kernel is operated, the resulting functions include the three-body breakup component as expressed by Eq. (\ref{eq:phi_asym}),  
and thus, the range of ${x}_1$ in such functions is not restricted to the range of the deuteron. 
As a result,  the cancellation as Eq. (\ref{eq:x3-y1}) is no more expected for higher order calculations.  
This is demonstrated by the fact that dashed curve in the figure, which denotes the source term $\omega_{\underline{\alpha}}^C(x_3)$  
including both of $V^C$ and $u^C$ contributions, remains to be non-negligible for a large value of $x_3$.

To include the long-range effect of $\omega_{\underline{\alpha}}^C(x_3)$ as much as possible, one needs to increase the value of $x_M$ much larger than 10 fm, which is the standard value in the $nd$ calculation.
This makes $pd$ calculations much harder than the $nd$ calculations.
Since Fig. \ref{fig:omega-xp} implies that the source function $\omega_{3,\underline{\alpha}}(x_3)$ is dominated by the 2NP contribution $\omega_{3,\underline{\alpha}}^S(x_3)$, we decide to include the effect of the long-range contribution partially by multiplying $\omega_{\underline{\alpha}}^C(x_3)$ by a cutoff factor
\begin{equation}
e^{-(x_3/R_C)^N}
\label{eq:cutoff}
\end{equation}
for higher order than the zeroth order in our iterative procedure of the MCF. 

%%%%%%%%%%%%%%%%%%%%%%%%%%%%%%%%%%%%%%%%%%%%
%%% Figure 3: Source functions \omega(x) %%%
%%%%%%%%%%%%%%%%%%%%%%%%%%%%%%%%%%%%%%%%%%%%

Validity of this procedure is examined in Table \ref{tab:av18-phase-rn}, where we compare some results of the $pd$ eigenphase shift and mixing angle parameters in convention defined in Ref. \cite{Hu95}, for a partial wave state with the total angular momentum and parity of $1/2^+$ calculated with the AV18 potential. 
In the table, the last column shows the results  by the Kohn variational principle (KVP) \cite{Vi01}. 
The rest columns do our calculations.
The first column denoted as WC0 shows results calculated by completely neglecting $\omega_{\underline{\alpha}}^C(x_3)$ in all order calculations of the MCF iteration scheme to solve the SSF equation. 
Calculations with keeping $\omega_{\underline{\alpha}}^C(x_3)$ only for the zeroth order without cutoff but neglecting $\omega_{\underline{\alpha}}^C(x_3)$ for higher order calculations are shown in the second column (WCn). 
In the third to fifth columns, we show results with the cutoff for non-zeroth order by choosing $N=4$ and $R_C$ = 4 fm, 6 fm, and 8 fm, respectively. 
A comparison of the numbers in the table indicates that effects due to the neglect of the long-range term $\omega_{\underline{\alpha}}^C(x_3)$ in the integral kernel may be an order of a few \% in the phase shift parameters, 
and suggests that the partial inclusion with the cutoff factor with $(N,R_C)$ = (4, 8 fm), e.g., may produce sufficient results.  
We remark that the imaginary part of the ${}^4$S$_{1/2}$ parameter reveals a rather slow convergence, which might affect scattering observables.
This point  will be discussed later.

%%%%%%%%%%%%%%%%%%%%%%%%%%%%%%%%%%%%%%%
%%% Table 1: phase shift parameters %%%
%%%%%%%%%%%%%%%%%%%%%%%%%%%%%%%%%%%%%%%
\begin{table*}[tb]
\caption{
\label{tab:av18-phase-rn}
The real and complex part of the $pd$ eigenphase shift and mixing angle parameters, which are given in degrees, for $J^{\pi}= 1/2^+$ state with the AV18 potential.
See the text for the meaning of the calculations.
}
\begin{ruledtabular}
\begin{tabular}{ccccccc}
 & WC0 & WCn & $N=4$ & $N=4$ & $N=4$ & KVP\footnote{Ref.\ \protect\cite{Vi01}}   \\
 &  &  & $R_C=4$ fm& $R_C=6$ fm & $R_C=8$ fm & \\
\hline
$E_p = 5.0$ MeV \\ 
${}^{4}$D$_{1/2}$ & (-5.33, 0.01) &(-5.44, -0.01) & (-5.44, -0.01)&(-5.45, 0.00) & (-5.45, 0.00) &(-5.43, 0.004)\\
${}^{4}$S$_{1/2}$ & (-42.4, 1.93) &(-42.4, 1.93)& (-41.8, 2.36) &(-41.8, 2.30) & (-41.8, 2.15) &(-41.8, 1.74)\\
$\eta_{1/2+}$     & (0.97, -0.04) &(1.01, -0.03) & (1.05, -0.03) &(1.05, -0.04) & (1.05, -0.04)&(1.05, -0.03)\\
\hline
$E_p = 10.0$ MeV \\
${}^{4}$D$_{1/2}$ & (-7.15, 0.24) &(-7.32, 0.21)& (-7.32, 0.22)&(-7.33, 0.22) & (-7.34, 0.22) &(-7.30, 0.24)\\
${}^{4}$S$_{1/2}$ & (-61.3, 11.6) &(-61.5, 11.5)& (-61.0, 12.4)&(-60.9, 11.9) & (-60.8, 11.9)& (-60.6, 11.7) \\
$\eta_{1/2+}$     & (0.96, 0.03) &(0.98, 0.05)& (1.02, 0.05)&(1.02, 0.04)  & (1.01, 0.04) &(1.01, 0.06) \\
\end{tabular}
\end{ruledtabular}
\end{table*}
%%%%%%%%%%%%%%%%%%%%%%%%%%%%%%%%%%%%%%%%%%%%%%%%%%%%%%%%%%%%%%%%%%%%

%%%%%%%%%%%%%%
%%% Sec. 4 %%%
%%%%%%%%%%%%%%
%%%%%%%%%%%%%%%%%%%%%%%%%%%%%%%%%%%%%%%%%%%%%%%%%%
\section{\label{sec:results} Numerical Results}
%%%%%%%%%%%%%%%%%%%%%%%%%%%%%%%%%%%%%%%%%%%%%%%%%%

In this section, we will present some numerical results by the formulation described in the previous sections. 
Technical details of introducing spins' degrees of freedom, 3NFs, etc., are given in Refs. \cite{Sa82,Sa83,Sa86,Is07}. 
As a standard 2NF model, we choose the AV18 potential \cite{Wi95}. 
Three-nucleon partial wave states, which the 2NF and 3NFs act, are restricted to those with total two-nucleon angular momenta $J \le 6$ for bound state calculations and $J \le 4$ for scattering calculations.
In scattering calculations, total $3N$ angular momentum is truncated at  $J_0 = 19/2$, while 3NF's are switched off for $3N$ states with $J_0>13/2$.

As described in the previous section, the Coulomb source term $\omega_{\underline{\alpha}}^C(x)$ in the SSF integral kernel is treated by multiplying with the cutoff factor Eq. (\ref{eq:cutoff}) for higher order in the MCF iteration. 
Comparisons of calculations performed by taking three different sets of $(N,R_C)$ in Table \ref{tab:av18-phase-rn} show that a satisfactory convergence is obtained with parameters of $(N,R_C)$ = (4, 8 fm) for elastic observables, 
and we thus proceed with these parameters referring to them simply as {\em  $pd$ calculations}. 
A convergence problem for three-body breakup observables will be discussed in a subsection below. 

Calculated binding energy of $^3$H ($^3$He) with the AV18 potential is 7.626 MeV (6.928 MeV), which is underbound by about 1 MeV compared to the empirical value of 8.482 MeV (7.718 MeV). 
It is well known that a 3NF that caused by the exchange of two pions among three nucleons (2$\pi$E-3NF) produces enough attraction to explain the empirical binding energy.
In this paper, we use a new version of the Brazil 2$\pi$E-3NF \cite{Is07b} with 
a dipole form factor of the cutoff mass parameter $\Lambda$,  $\left(\frac{\Lambda^2-m_\pi^2}{\Lambda^2+\bm{q}^2}\right)^2$ for the $\pi NN$ vertex (BR$_{\Lambda}$).
In a combination with the AV18 2NP, $\Lambda$ is chosen to be 660 MeV (AV18+BR$_{660}$) to give  the binding energy 8.492 MeV (7.763 MeV) for $^3$H ($^3$He).

%%%%%%%%%%%%%%%%%%%%%%%%%%%%%%%%%%%%%%%%%%%%%%%%%%%%%%%%%%%%%
\subsection{Differential cross section in elastic scattering}
\begin{figure}[tb]
\includegraphics[width=0.80\columnwidth,angle=0]{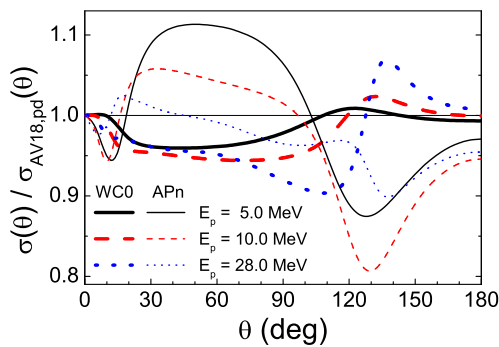}
\caption{(Color online) 
Differential cross section of $pd$ elastic scattering normalized by the $pd$ calculation with the AV18 potential. 
The bold curves represent the AV18-WC0 calculations, and the thin curves the AV18-APn calculations.
The solid curves denote calculations at $E_p$ = 5.0 MeV, the dashed curves $E_p$ = 10.0 MeV, and the dotted curves $E_p$ = 28.0 MeV. 
\label{fig:pd-approx-ratio}}
\end{figure}

%%%%%%%%%%%%%%%%%%%%%%%%%%%%%%%%%%%%%%%%%%%%%%%%%%%%%%%%%%%%%
First, we compare calculations approximately including the $pp$ Coulomb force effects with those of the $pd$ calculations for the differential cross section $\sigma(\theta)$ of the $pd$ elastic scattering, where $\theta$ is the scattering angle in the c.m. system. 
We take two approximate calculations: one is the WC0 calculation, which is presented in the previous section.
The other one, which will be denoted as APn,  is an approximate calculation, in which the scattering amplitude due to the short-range 2NF 
is replaced by a corresponding $nd$ scattering amplitude \cite{Do82}.
It is expected that the WC0 calculations are better approximation at lower energies since breakup effects are smaller.
On the other hand, the APn calculations are expected to be better for higher energies. 
Fig.\ \ref{fig:pd-approx-ratio}, where WC0 (bold curves) and APn calculations (thin curves) of differential cross sections normalized by those with the $pd$ calculations at $E_p$ = 5.0, 10.0, and 28.0 MeV are plotted, looks to exhibit roughly these tendencies.   
It is remarkable that deviations of the APn calculations are rather large, about 10 \% even at backward angles.

%%%%%%%%%%%%%%%%%%%%%%%%%%%%%%%%%%%%%%%%%%%%%%%%%%%%%%%%%%%%%%%%%%%%
%%%%%%%%%%%%%%%%%%%%%%%%%%%%%%%%%%%%%%%%%%%%%%%%%%%%%%%%%%%%%%%%%%%%%%%

\begin{figure}[tb]
\includegraphics[width=0.70\columnwidth,angle=0]{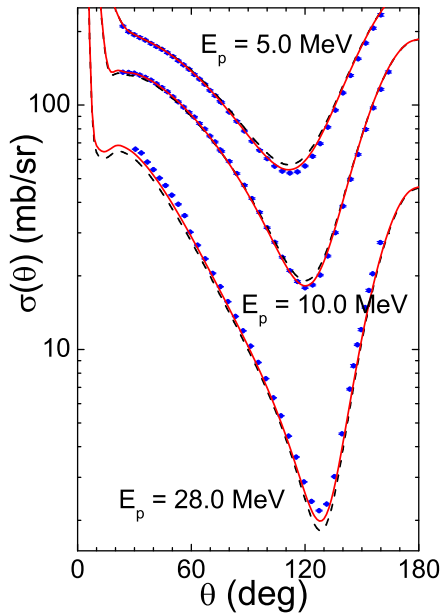}
\caption{(Color online) 
Differential cross sections of $pd$ elastic scattering at $E_p$ = 5.0, 10.0, and 28.0 MeV. 
The dashed curves denote calculations with the AV18 potential and the solid curves those with the AV18+BR$_{660}$ potential.
Experimental data are from Ref. \protect\cite{Sa94} for $E_p$ = 5.0 MeV and 10.0 MeV, and Ref.  \protect\cite{Ha84} for $E_p$ = 28.0 MeV. 
\label{fig:pd-crs}}
\end{figure}

\begin{figure}[tb]
\includegraphics[width=0.80\columnwidth,angle=0]{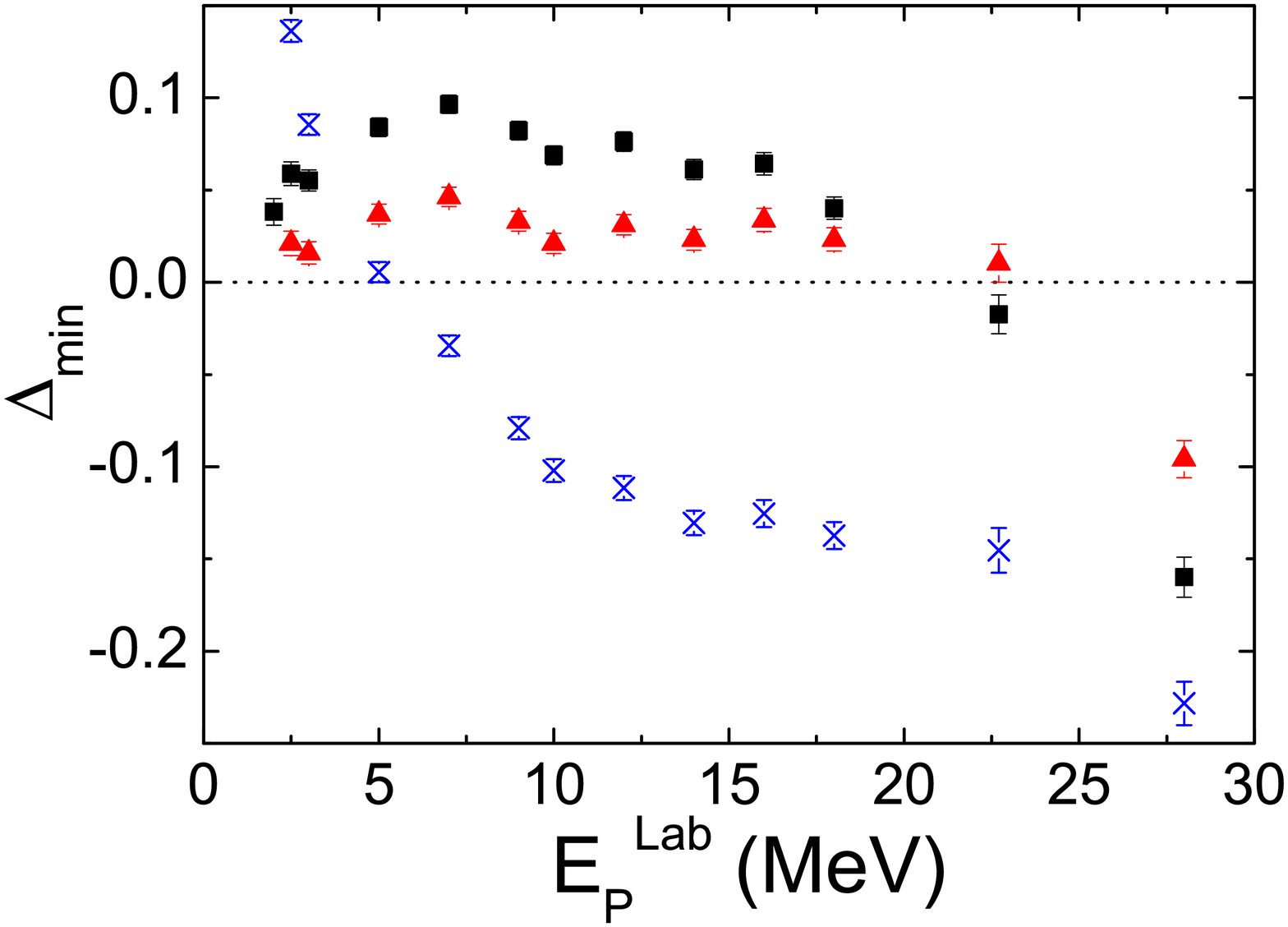}
\caption{(Color online)
Discrepancy of the $pd$ differential cross section minimum $\Delta_{min}$ defined in Eq. (\protect\ref{eq:dcs_min}) using the experimental data \protect\cite{Sa94,Ha84,Gr83}.
The solid squares denote the $pd$ calculations with the AV18 potential, the triangles those with the AV18 + BR$_{660}$ potential, and the crosses the APn calculations with the AV18 potential.
\label{fig:pd-crs-min} }
\end{figure}

In Fig. \ref{fig:pd-crs}, $pd$ calculations for $\sigma(\theta)$ of the $pd$ elastic scattering at $E_p$ = 5.0 MeV, 10.0 MeV, and 28.0 MeV are plotted together with experimental data \cite{Sa94,Ha84}. 
Comparing the calculations with the AV18 potential (the dashed curves) to the experimental data at  $\theta\sim 120^\circ$, where $\sigma(\theta)$ takes the minimum, one finds that the calculations overestimate the data at lower energies and underestimate at higher energies. 
The introduction of the 2$\pi$E-3NF as shown by the solid curves reduces almost all of the discrepancies at lower energies.
This systematic difference between the 2NF calculations and the data, which is referred to as \textquotedblleft {\em Sagara discrepancy}\textquotedblright, was pointed out in Ref. \cite{Sa94} using the APn calculations. 
To study this discrepancy in detail, we plot a relative discrepancy between the data 
\cite{Sa94,Ha84,Gr83} and calculations defined by
\begin{equation}
\Delta_{min}
=\frac{\sigma^{calc}(\theta_{min})-\sigma^{exp}(\theta_{min})}{\sigma^{exp}(\theta_{min})}
\label{eq:dcs_min}
\end{equation}
in Fig. \ref{fig:pd-crs-min}, where $\theta_{min}$ is the scattering angle where the differential cross section takes the minimum. 
The $pd$ calculation shows that a systematic discrepancy still remains when effects of the Coulomb force are treated properly, but with shifting transition energy from the overestimation to the underestimation to a higher energy of about $E_p$ = 20 MeV as compared to that by the APn calculation, about 5 MeV. 
This tendency is consistent with the results reported in Refs. \cite{Ki01,Al02}.

%%%%%%%%%%%%%%%%%%%%%%%%%%%%%%%%%%%%%%%%%%%%%%%%%%%%%%%%%%%%%%%%%%%%
%%%%%%%%%%%%%%%%%%%%%%%%%%%%%%%%%%%%%%%%%%%%%%%%%%%%%%%%%%%%%%%%%%%%%%%

%%%%%%%%%%%%%%%%%%%%%%%%%%%%%%%%%%%%%%%%%%%%%%%%%%%%%%%%%%%%%%%%%%%%
%%%%%%%%%%%%%%%%%%%%%%%%%%%%%%%%%%%%%%%%%%%%%%%%%%%%%%%%%%%%%%%%%%%%%

%%%%%%%%%%%%%%%%%%%%%%%%%%%%%%%%%%%%%%%%%%%%%%%%%
\subsection{Phenomenological three-nucleon force}
\begin{figure}[tb]
\includegraphics[width=0.95\columnwidth,angle=0]{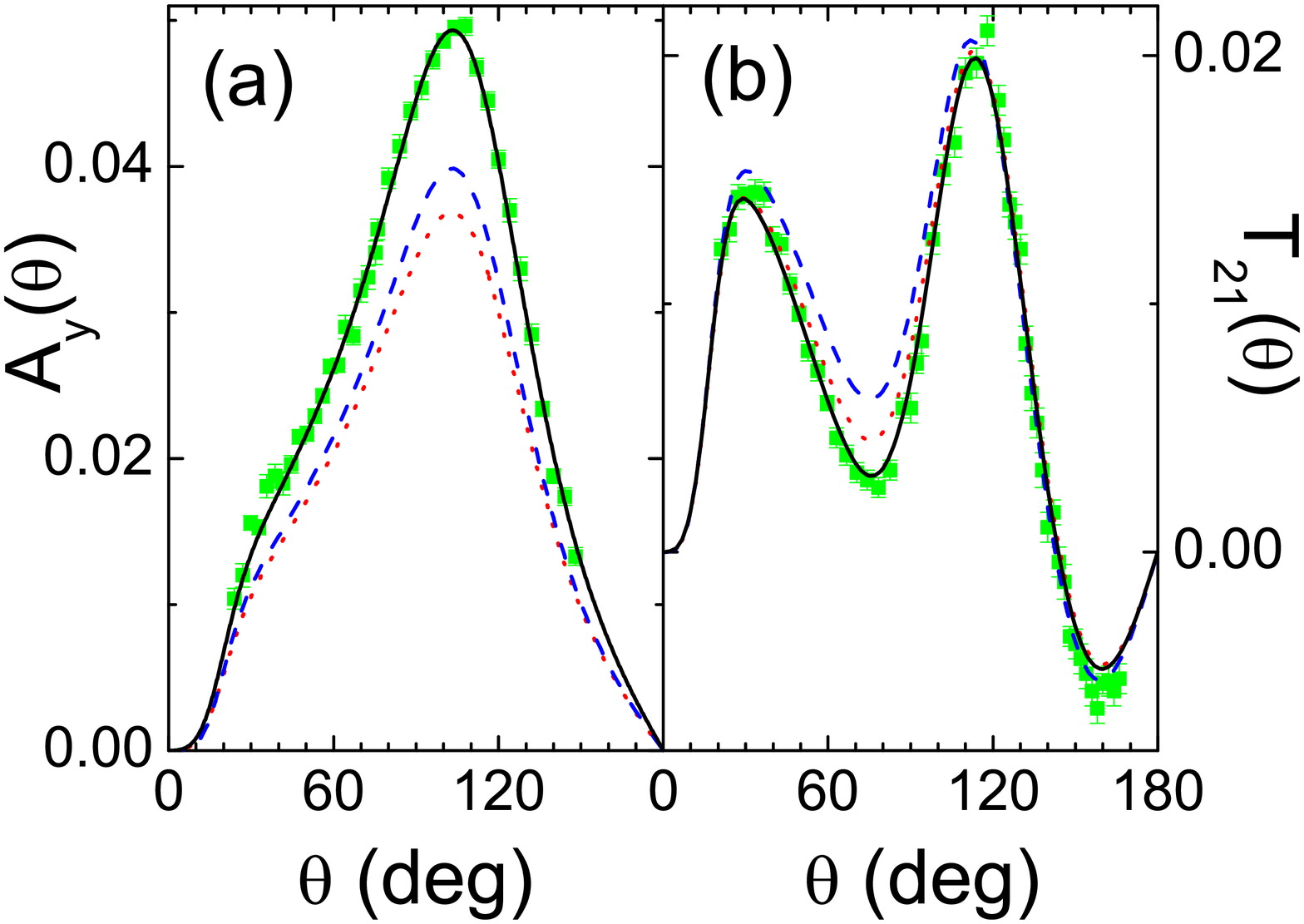}
\caption{(Color online)
(a) Proton vector analyzing power $A_y(\theta)$ and (b) deuteron tensor analyzing power $T_{21}(\theta)$ of $pd$ elastic scattering at $E_p$ = 3.0 MeV. 
The dotted curves denote calculations with the AV18 potential, the dashed curves those with the AV18+BR$_{660}$ potential, and the solid curves those with the AV18+BR$_{800}$+V$^{phe}$ potential.
Experimental data are from Refs. \protect\cite{Sa94,Sh95}.
\label{fig:ay-t21-3mev}}
\end{figure}

%%%%%%%%%%%%%%%%%%%%%%%%%%%%%%%%%%%%%%%%%%%%%%%%%
In Ref. \cite{Is03b}, it is pointed out that the introduction of the 2$\pi$E-3NF causes an undesirable effect to the tensor analyzing power $T_{21}(\theta)$ of the $pd$ elastic scattering at energies below the TBT.
Also, the 2$\pi$E-3NF is known to give little effect on the vector analyzing power $A_y(\theta)$, for which there exists rather large discrepancy between experimental data and calculations (\textquotedblleft {\em $A_y$ puzzle}\textquotedblright). 
These facts, which are also demonstrated in Fig. \ref{fig:ay-t21-3mev}, suggest that 
the 2$\pi$E-3NF is insufficient to comprise a Nuclear Hamiltonian in addition to the realistic 2NF.
Since no possible mechanism to produce additional 3NF to remedy above defects is established, at the moment, a phenomenological 3NF model is introduced \cite{Is07c}, which has a form that typical components in 2NP:  central, tensor, and spin-orbit components,  are modified in the presence of third nucleon.
The explicit form of the 3NF is  
\begin{eqnarray}
V^{phe} &=&  \sum_{i<j} e^{-\left(\frac{r_{ik}}{r_G}\right)^2 - \left(\frac{r_{jk}}{r_G}\right)^2 } \left[ V_0 + V_T S_T(ij) \hat{P}_{11} \right] 
\cr
&+& V_{ls} e^{ -\alpha_{ls} \rho }  \sum_{i<j}
  \left[ \bm{\ell}_{ij}\cdot\left(\bm{S}_i + \bm{S}_j \right) \right] \hat{P}_{11},
\label{eq:phe-3np}
\end{eqnarray}
where 
$S_{T}(ij)$ is the tensor operator acting between the nucleon pair $ij$, 
$\hat{P}_{11}$ is the projection operator to the spin and isospin triplet state of the pair $ij$, and $\rho=\frac23\left(r_{12}^2 + r_{23}^2 + r_{31}^2 \right)$.
The range parameter $r_G$ was taken to be 1.0 fm, and $\alpha_{ls}$ to be 1.5 fm$^{-1}$.
In Ref. \cite{Is07c}, the strength parameters, $V_{0}$, $V_{T}$, and $V_{ls}$, are determined in the following manner:
We choose the combination of the AV18 potential for 2NF and a former version of Brazil model \cite{Co83} with $\Lambda=$ 800 MeV for the 2$\pi$E-3NF as a starting interaction, which makes the triton overbound by about 1 MeV.
The parameters are decided to reproduce the following observables:
the triton binding energy, vector analyzing power $A_y(\theta)$ and tensor analyzing power $T_{21}(\theta)$ in $pd$ scattering at $E_p$ = 3.0 MeV.  
The results are $V_{0} = 25 $ MeV, $V_{T} = -40$ MeV, and $V_{ls} = -16$ MeV.

In the present work,  we will use a new version of Brazil 3NF \cite{Is07b}, which is more attractive in the $3N$ bound states than the earlier version \cite{Co83}. 
We thus need to retune the value of $V_{0}$ to be 36 MeV, but without changing the values of $V_{T}$ and $V_{ls}$.
Calculated binding energy with this set of potentials (AV18+BR$_{800}$+V$^{phe}$) is 8.482 MeV  (7.757 MeV) for $^3$H ($^3$He), and the results for $A_y(\theta)$ and $T_{21}(\theta)$ at $E_p=3.0$ MeV is shown by solid curves in  Fig. \ref{fig:ay-t21-3mev}.

%%%%%%%%%%%%%%%%%%%%%%%%%%%%%%%%%%%%%%%%%%%%%%%%%%%%%%%%%%%%%%%%%%%%
%%%%%%%%%%%%%%%%%%%%%%%%%%%%%%%%%%%%%%%%%%%%%%%%%%%%%%%%%%%%%%%%%%%%%%%

%%%%%%%%%%%%%%%%%%%%%%%%%%%%%%%%%%%%%%%%%%%%%%%%%%%%%%%%%%%
\subsection{Polarization observables in elastic scattering}
\begin{figure*}[tb]
\includegraphics[width=1.5\columnwidth,angle=0]{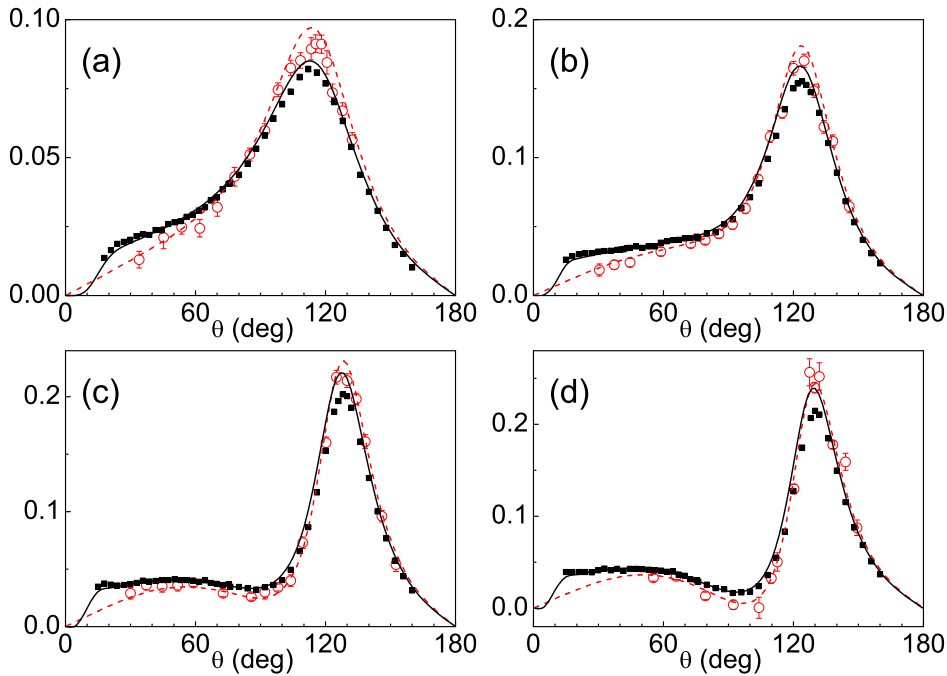}
\caption{(Color online) 
Nucleon vector analyzing power $A_y(\theta)$ of $pd$ and $nd$ elastic scattering at $E_N$ = (a) 5.0  MeV, (b) 10.0 MeV, (c) 14.0 MeV, and (d) 16.0 MeV. 
Solid (dashed) curves denote $pd$ ($nd$) calculations with the AV18+BR$_{800}$+V$^{phe}$ potential. 
Experimental data are from Refs. \protect\cite{Sa94,Sh95} for $pd$ (solid squares) and Refs.  \protect\cite{Ho87,Fu99} for $nd$ (open circles).
\label{fig:ay-e-dep}}
\end{figure*}

%%%%%%%%%%%%%%%%%%%%%%%%%%%%%%%%%%%%%%%%%%%%%%%%%%%%%%%%%%%
In Ref. \cite{Is07c}, it is shown that the use of the phenomenological 3NF together with the AV18+BR$_{800}$, which is tuned to reproduce the $3N$ binding energy; $A_y(\theta)$ and $T_{21}(\theta)$ at $E_p$ = 3.0 MeV, is also successful in describing the neutron vector analyzing power $A_y(\theta)$ of the $nd$ scattering at higher energies. 
In Fig. \ref{fig:ay-e-dep}, calculations of $A_y(\theta)$ of the $pd$ and $nd$ scattering with the AV18+BR$_{800}$+$V^{phe}$ potentials are compared with experimental data at some energies above the TBT \cite{Sa94,Sh95,Ho87,Fu99}.  
While the calculations of the $nd$-$A_y(\theta)$ agree with the experimental data in similar manner as in Ref. \cite{Is07c}, those of the $pd$-$A_y(\theta)$ overestimate the data at the maximum region $\theta \sim 130^\circ$ as the energy increases. 
In another aspect, the calculated difference between the $nd$- and the $pd$-$A_y(\theta)$ at the maximum region is decreasing as the energy increases, which is contradictory to the tendency of the experimental data.

%%%%%%%%%%%%%%%%%%%%%%%%%%%%%%%%%%%%%%%%%%%%%%%%%%%%%%%%%%%%%%%%%%%%
%%%%%%%%%%%%%%%%%%%%%%%%%%%%%%%%%%%%%%%%%%%%%%%%%%%%%%%%%%%%%%%%%%%%%%%

\begin{figure}[tb]
\includegraphics[width=0.80\columnwidth,angle=0]{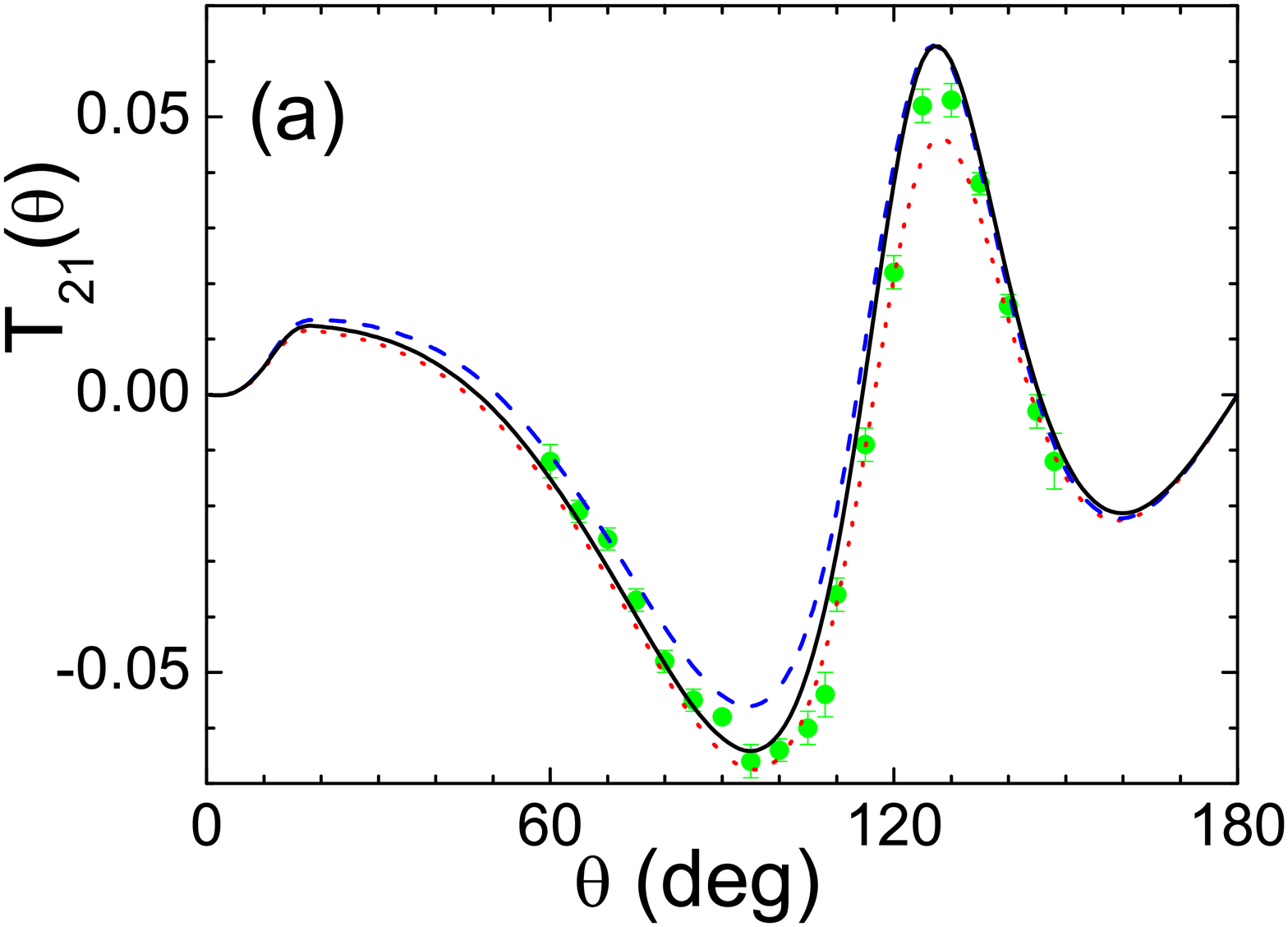}
\includegraphics[width=0.80\columnwidth,angle=0]{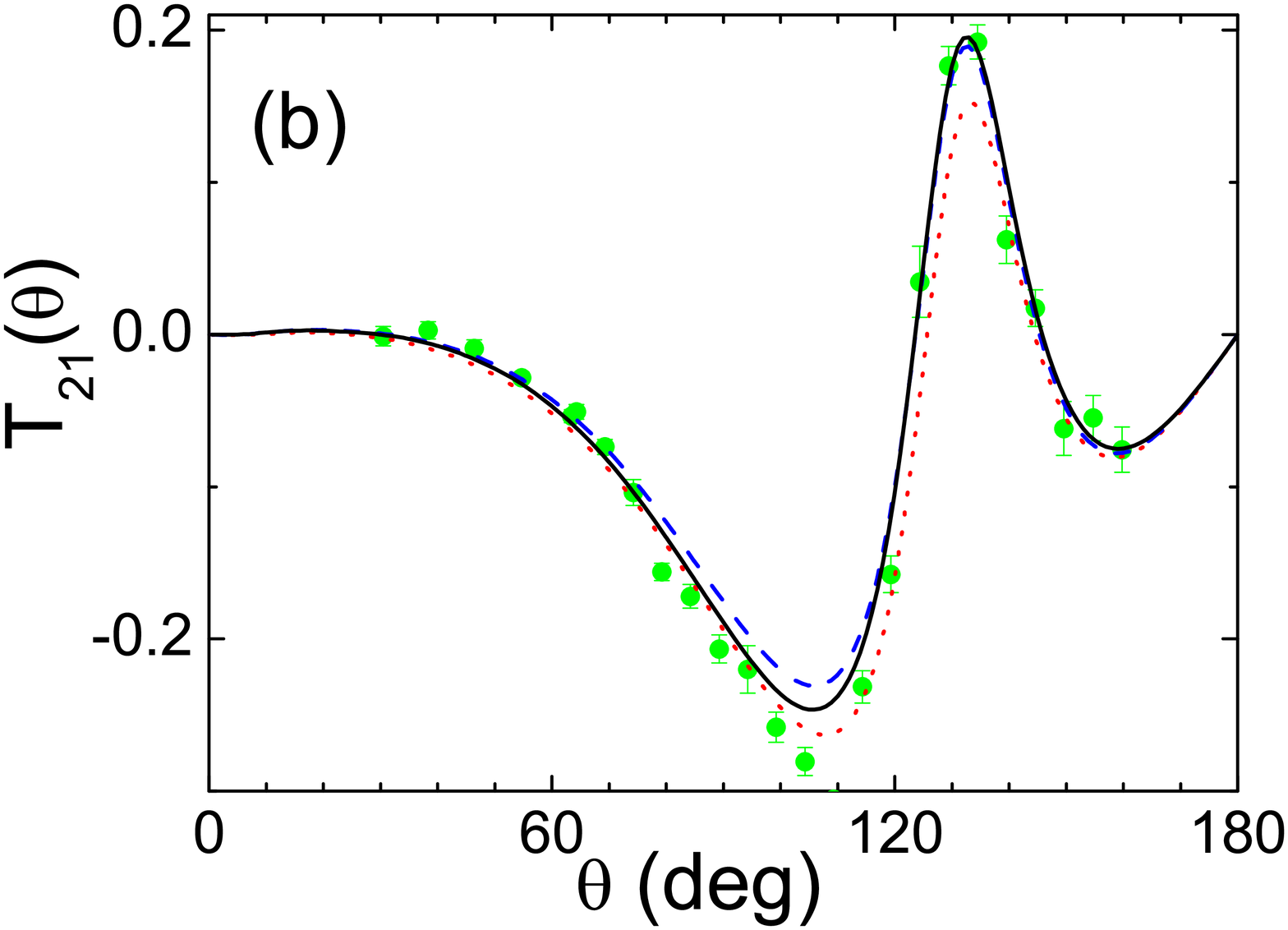}
\caption{(Color online)
Deuteron tensor analyzing power $T_{21}(\theta)$ of $pd$ elastic scattering at $E_p$ = (a) 10.0 MeV and (b) 28.0 MeV (or equivalently $E_d$ = 20.0 MeV and 56.0 MeV, respectively, in a deuteron incident scattering). 
The dotted curves denote calculations with the AV18 potential, the dashed curves those with the AV18+BR$_{660}$ potential, and the solid curves those with the AV18+BR$_{800}$+V$^{phe}$ potential.
Experimental data are from Ref. \protect\cite{Sa83ex} for $E_p$ = 10.0 MeV and Ref. \protect\cite{Ha84} for $E_p$ = 28.0 MeV.
\label{fig:t21}}
\end{figure}

In Fig. \ref{fig:t21}, calculations of the deuteron tensor analyzing power $T_{21}(\theta)$ of the $pd$ scattering at $E_p$ = 10.0 MeV and 28.0 MeV are compared with experimental data \cite{Sa83ex,Ha84}. 
As in the case at the low energy, the introduction of the 2$\pi$E-3NF shifts the calculations to wrong direction from the experimental data around $\theta=90^\circ$, and the phenomenological 3NF works to reproduce the data.
Another interesting feature appears at $\theta\sim130^\circ$, where $T_{21}(\theta)$ takes the maximum as follows: while the calculations with the AV18 potential deviate from the data, the introduction of the both 3NFs results the almost the same and remedy the discrepancy well. 
These suggest that $T_{21}(\theta \sim 90^\circ)$ is sensitive to tensor components of nuclear forces and  $T_{21}(\theta \sim 130^\circ)$ to central components.

\begin{figure}[tb]
\includegraphics[width=0.80\columnwidth,angle=0]{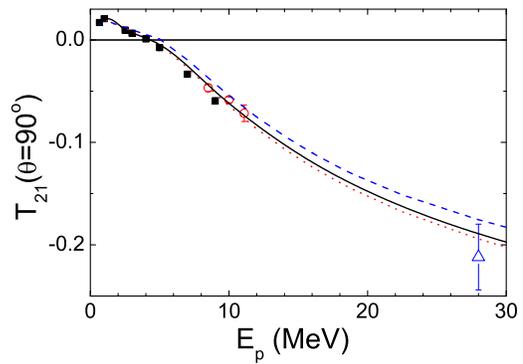}
\caption{(Color online)
Energy dependence of deuteron tensor analyzing power $T_{21}(\theta=90^\circ)$ of $pd$ elastic scattering. 
The dotted curve denotes calculation with the AV18 potential, the dashed curve one with the AV18+BR$_{660}$ potential, and the solid curve one with the AV18+BR$_{800}$+V$^{phe}$ potential. 
Experimental data are from Ref. \protect\cite{Sh95} (squares), Ref. \protect\cite{Sa83ex} (circles), and Ref. \protect\cite{Ha84} (triangle).
\label{fig:21-90-edep}}
\end{figure}

In Fig. \ref{fig:21-90-edep}, energy dependence of the deuteron tensor analyzing power $T_{21}(\theta)$ at $\theta=90^\circ$ in the $pd$ elastic scattering for calculations with the AV18, AV18+BR$_{660}$, and AV18+BR$_{800}$+V$^{phe}$ potentials is shown in comparing with available data \cite{Sa94,Sh95,Sa83ex,Ha84}.
The figure shows that the introduction of the phenomenological 3NF is still consistent with data at higher energies. 
It is interesting if further  $T_{21}(\theta)$ data at $E_p$ = 20 MeV to 30 MeV, where experimental data are missing, are consistent with the calculation or not.

%%%%%%%%%%%%%%%%%%%%%%%%%%%%%%%%%%%%%%%%%%%%%%%%%%%%%%%%%%%%%%%%%%%%
%%%%%%%%%%%%%%%%%%%%%%%%%%%%%%%%%%%%%%%%%%%%%%%%%%%%%%%%%%%%%%%%%%%%%%%

%%%%%%%%%%%%%%%%%%%%%%%%%%%%%%%%%%%%%%%%%%%%%%%%%%%%%%%%%%%%%%%%%%%%
%%%%%%%%%%%%%%%%%%%%%%%%%%%%%%%%%%%%%%%%%%%%%%%%%%%%%%%%%%%%%%%%%%%%%%%

\begin{figure}[tb]
\includegraphics[width=0.80\columnwidth,angle=0]{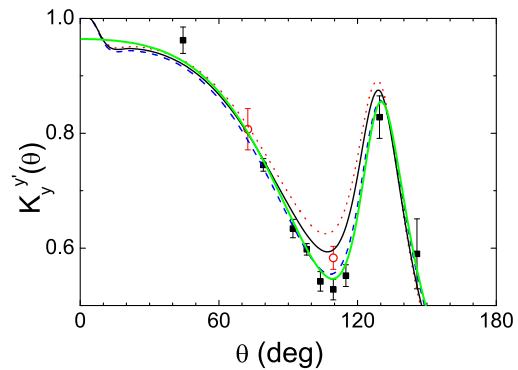}
\caption{(Color online) 
Polarization-transfer coefficient $K_y^{y^\prime}(\theta)$ of $pd$ and $nd$ elastic scattering at $E_N$ = 19.0 MeV. 
The dotted curve denotes the $pd$ calculation with the AV18 potential,  
the dashed curve the $pd$ one with the AV18+BR$_{660}$ potential,
the solid curve the $pd$ one with the AV18+BR$_{800}$+V$^{phe}$ potential,  
and the bold curve the $nd$ one with the AV18+BR$_{800}$+V$^{phe}$ potential. 
Experimental data are from Ref. \protect\cite{Sy98} for $pd$ (solid squares) and from Ref. \protect\cite{He98} for $nd$ scattering (open circles).
\label{fig:kyys-19mev}}
\end{figure}

Polarization-transfer coefficients are another interesting observables, which are sensitive to spin-dependent interactions. 
In Fig. \ref{fig:kyys-19mev}, the polarization-transfer coefficient $K_y^{y^\prime}(\theta)$ of  $pd$ and $nd$ elastic scattering at $E_N$ = 19.0 MeV are compared with experimental data \cite{Sy98,He98}. 
One interesting point, which has been already remarked in Ref. \cite{Ki01b}, is that the Coulomb force effect in the calculation is opposite to that in the data at $\theta\sim110^\circ$. 
In addition, the figure shows that the experimental data indicates that the AV18+BR$_{660}$ potential is favored than the AV18+BR$_{800}$+V$^{phe}$, implying this observable may be useful in distinguishing various 3NF models that reproduce other observables equally.

%%%%%%%%%%%%%%%%%%%%%%%%%%%%%%%%%%%%%%%%%%%%%%%%%%%%%%%%%%%%%%%%%%%%
%%%%%%%%%%%%%%%%%%%%%%%%%%%%%%%%%%%%%%%%%%%%%%%%%%%%%%%%%%%%%%%%%%%%%%%

%%%%%%%%%%%%%%%%%%%%%%%%%%%%%%%%%%
\subsection{Breakup cross section}
%%%%%%%%%%%%%%%%%%%%%%%%%%%%%%%%%%

Finally, we will show some results for differential cross sections of kinematically complete three-body breakup reactions, $d(p,pp)n$ and $d(n,nn)p$, which are characterized by configurations of three particles in the final state. 
Here, we will discuss four different kinematical conditions that include the following typical configurations, whose experimental data at $E_N$ = 13.0 MeV are available for the $pd$-breakup in Ref. \cite{Ra91} and for the $nd$-breakup in Refs. \cite{St89,Se05}:

{(a)}
Collinear (COL) configuration, in which three nucleons align on a line with the unobserved nucleon being at rest in the c.m. system;  

{(b)}
Final state interaction (FSI) configuration, in which the relative energy between the unobserved nucleon and one of the observed nucleon is zero;

{(c)}
Space star (SST) configuration, in which three nucleons have equal energies and interparticle angles of 120$^{\circ}$ in the c.m. system, and the plane spanned by the three nucleons is orthogonal to the beam axis;

{(d)}
Quasi-free scattering (QFS) configuration, in which the unobserved nucleon is at rest in the laboratory system. 

\begin{figure*}[tb]
\includegraphics[width=0.80\columnwidth,angle=0]{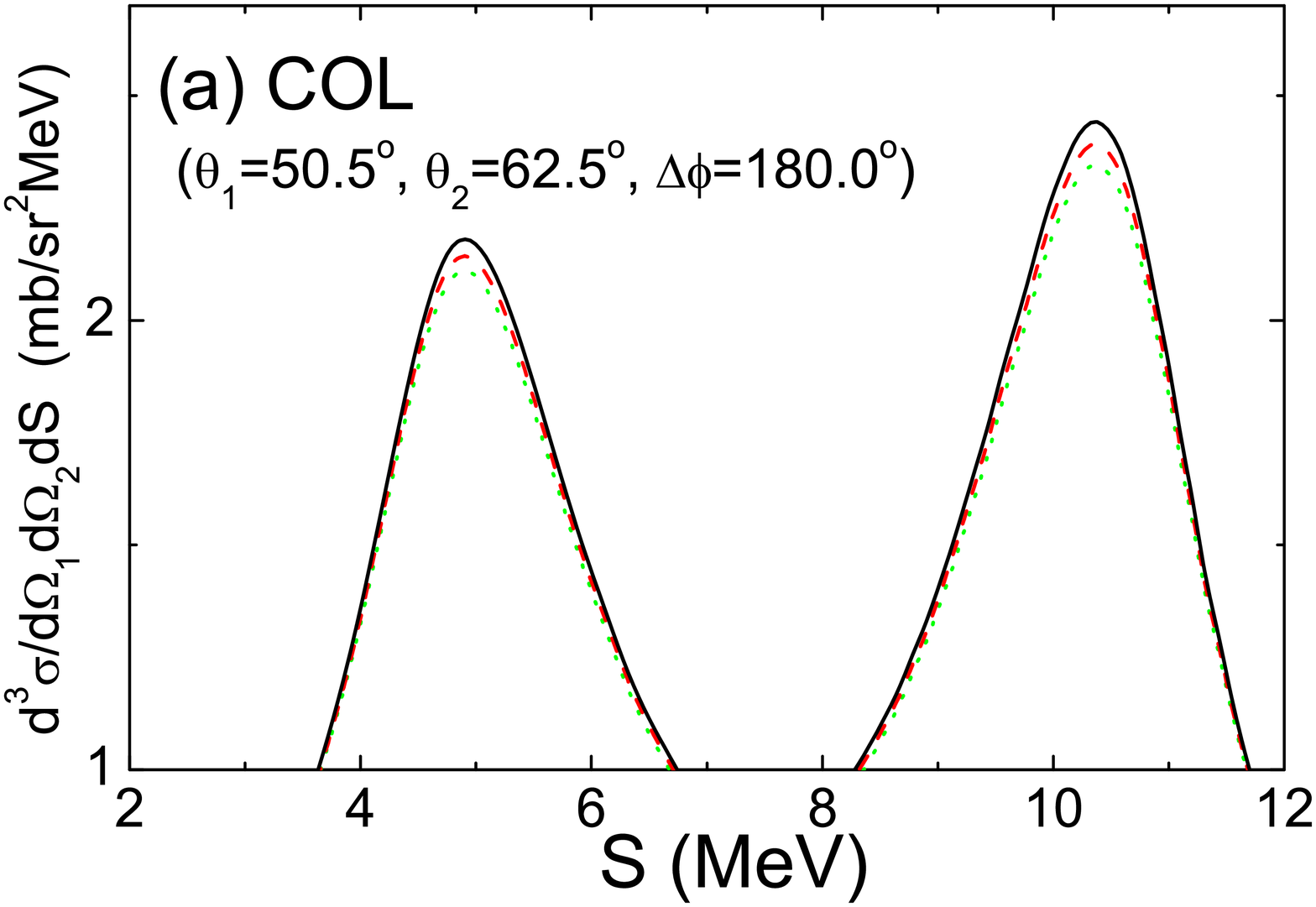}  %0.45 or 
\includegraphics[width=0.80\columnwidth,angle=0]{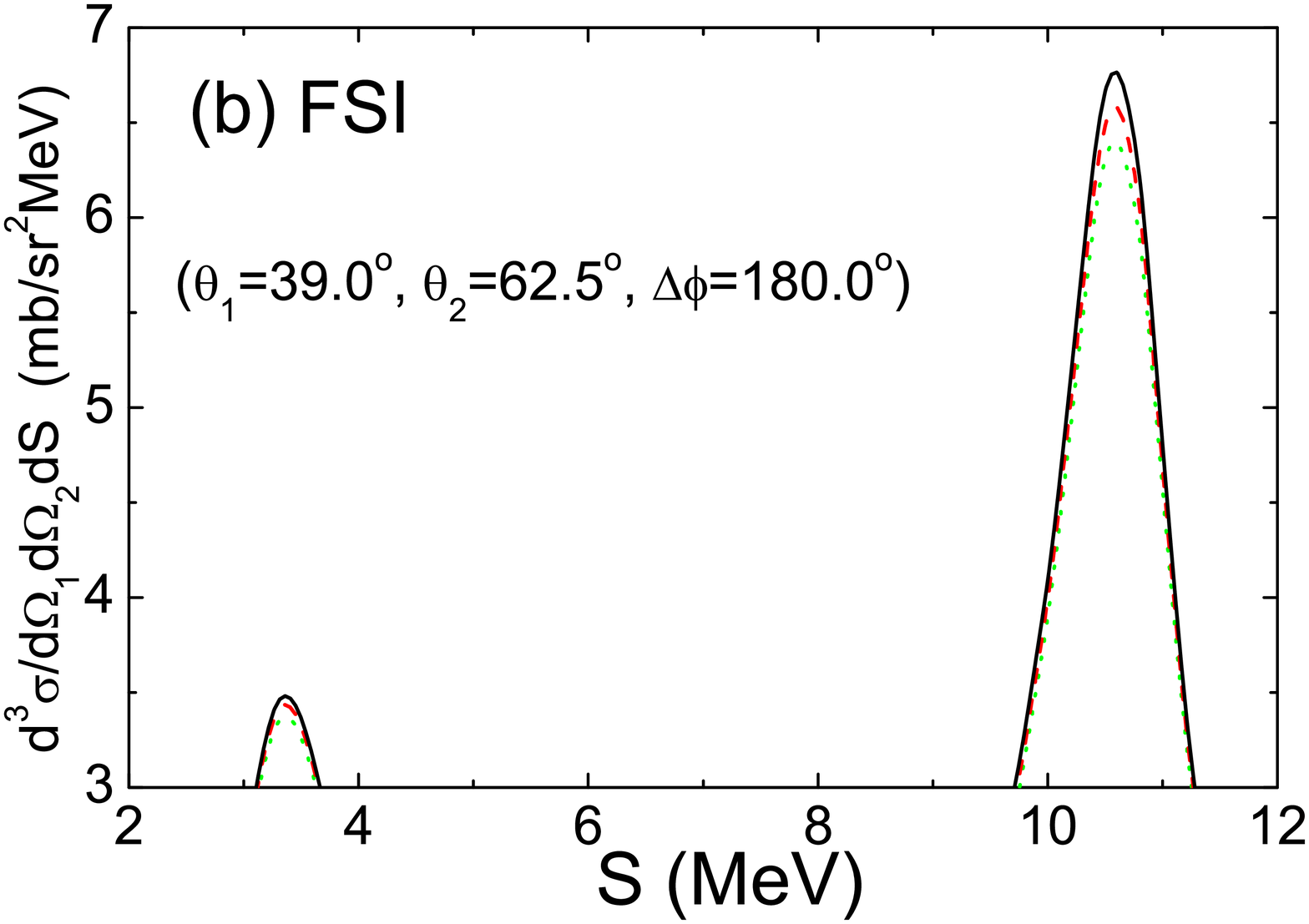}
\includegraphics[width=0.80\columnwidth,angle=0]{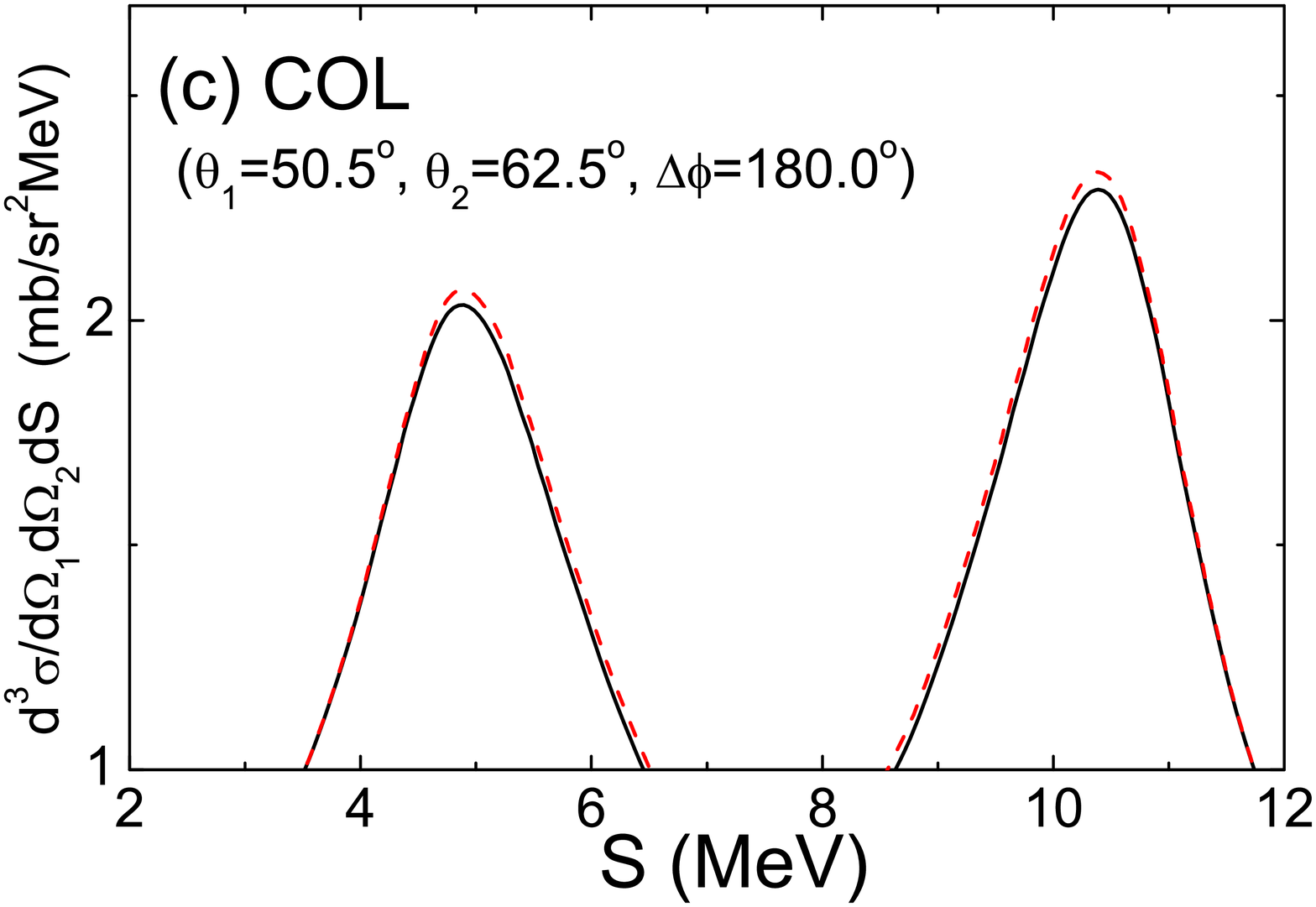}
\includegraphics[width=0.80\columnwidth,angle=0]{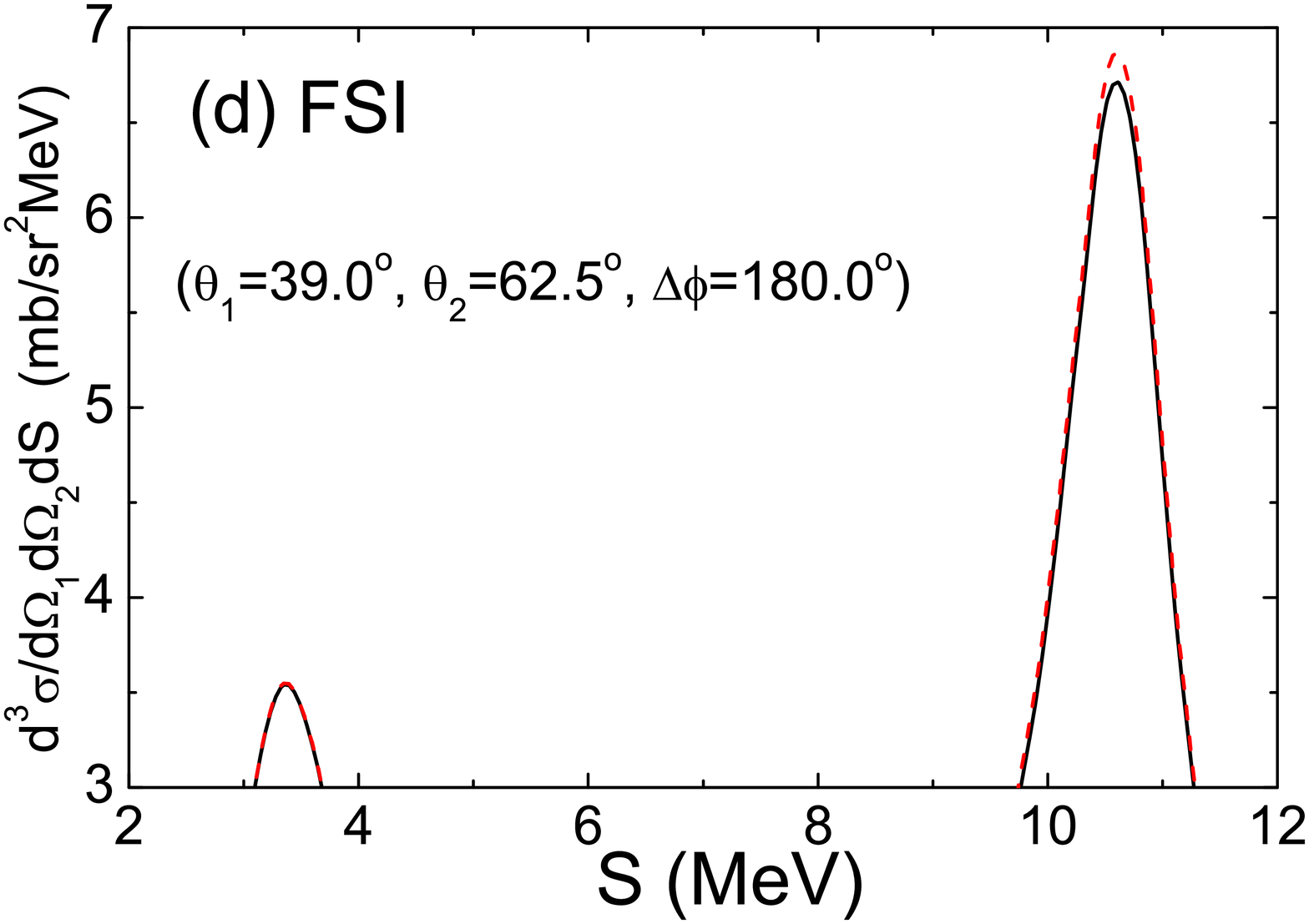}
\caption{(Color online) 
Differential cross sections of $pd$ breakup reactions (a) for the COL configuration and (b) for the FSI configuration;
those of $nd$ breakup reactions (c) for the COL configuration and (d) for the FSI configuration,  at $E_N$ = 13.0 MeV.
In (a) and (b), dotted curves denote calculations with $(N,R_C)=$ (4, 4 fm) for the AV18 potential, 
dashed curves $(N,R_C)=$ (4, 6 fm), and
solid curves $(N,R_C)=$ (4, 8 fm). 
In (c) and (d), solid curves denote the calculations with the AV14 potential and
dashed curves the AV14' potential.
\label{fig:pd-13mev-check}}
\end{figure*}

First, we have checked a convergence of the breakup cross sections with respect to the cutoff procedure of the long-range Coulomb force effect with Eq. (\ref{eq:cutoff}). 
Calculations with three parameter sets shown in Table \ref{tab:av18-phase-rn}, namely $(N, R_C) =$ (4, 4 fm), (4, 6 fm), and (4, 8 fm), for the SST and the QFS configurations agree with one another excellently, however those for the COL and the FSI configurations do in part as shown in Figs. \ref{fig:pd-13mev-check} (a) and (b). 

The visible deviations in Figs. \ref{fig:pd-13mev-check} (a) and (b) appear at a kinematical condition where the relative $pn$ energy is small, {\textit i.e.}, $E_{pn} < 0.5$ MeV, which might be caused by a small change in the $pp$ interaction due to our Coulomb treatment.  
To check this peculiar behavior, we have investigated the dependence of the $pn$-FSI cross sections in the $nd$ breakup reaction on the neutron-neutron ($nn$) interaction using two different 2NP models: a charge independent 2NP, Argonne V$_{14}$ (AV14) \cite{Wi84}, in which the $nn$ force is equal to the $pn$ force in the ${}^1$S$_{0}$ state, and its modified version (AV14') made in Refs. \cite{Wu90,Wu93} by considering an charge-dependent potential to distinguish the $nn$ force from the $pn$ force.
Results are shown in Figs. \ref{fig:pd-13mev-check} (c) and (d), which demonstrate that a change in a $nn$ force actually results in non-negligible effects for the $pn$-FSI cross sections.

\begin{figure}[tb]
\includegraphics[width=0.90\columnwidth,angle=0]{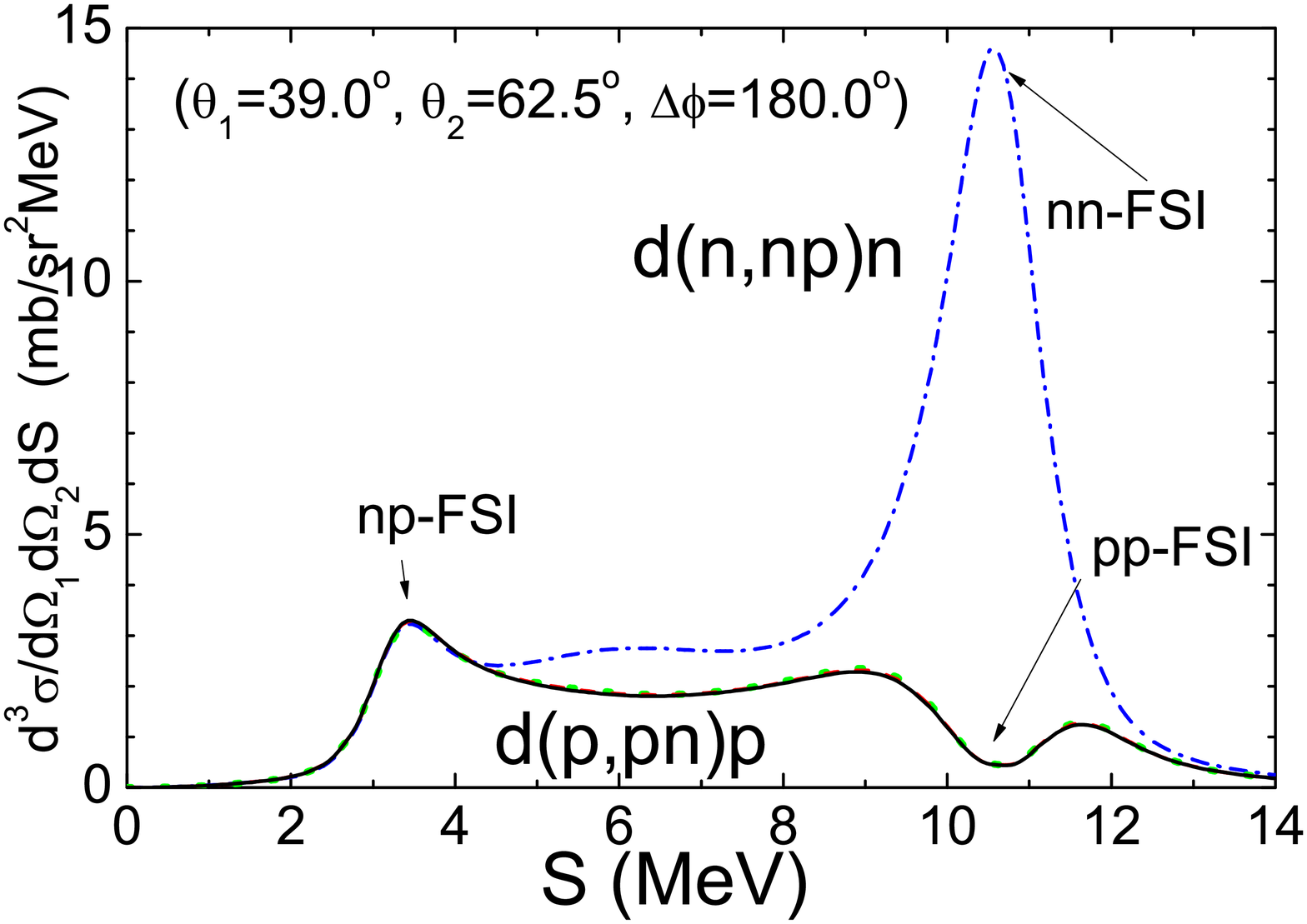}
\caption{(Color online)
Differential cross sections of $d(p,pn)p$ and $d(n,np)n$ reactions for the FSI configuration at $E_N$ = 13.0 MeV with the AV18 potential. 
Dotted curves denote calculations for the $pd$ breakup reaction with $(N,R_C)=$ (4, 4 fm), 
dashed curves $(N,R_C)=$ (4, 6 fm), and
solid curves $(N,R_C)=$ (4, 8 fm).
Dot-dashed curve denotes the cross section for the $nd$ breakup reaction.
\label{fig:dn-np-fsi-13mev}}
\end{figure}

Next, as a reference, cross sections of the FSI configuration for $d(p,pn)p$ and $d(n,np)n$ reactions at $E_N = 13.0$ MeV are plotted in Fig. \ref{fig:dn-np-fsi-13mev}.
In this configuration, the $pn$-FSI occurs around $S = 3$ MeV and the $pp$-FSI or $nn$-FSI does around $S= 11$ MeV. 
The $pp$-FSI cross sections are suppressed by the $pp$ Coulomb force compared to the $nn$-FSI cross sections, but not completely. 
This may suggest that we need to improve the Coulomb cutoff procedure possibly by extending the range of the cutoff function to treat the $pp$-FSI more correctly.

From these considerations, we conclude that our calculations successfully converge for the most of breakup configurations possibly except for a limited region with relative energy of two nucleons being close to zero.

\begin{figure*}[tb]
\includegraphics[width=0.80\columnwidth,angle=0]{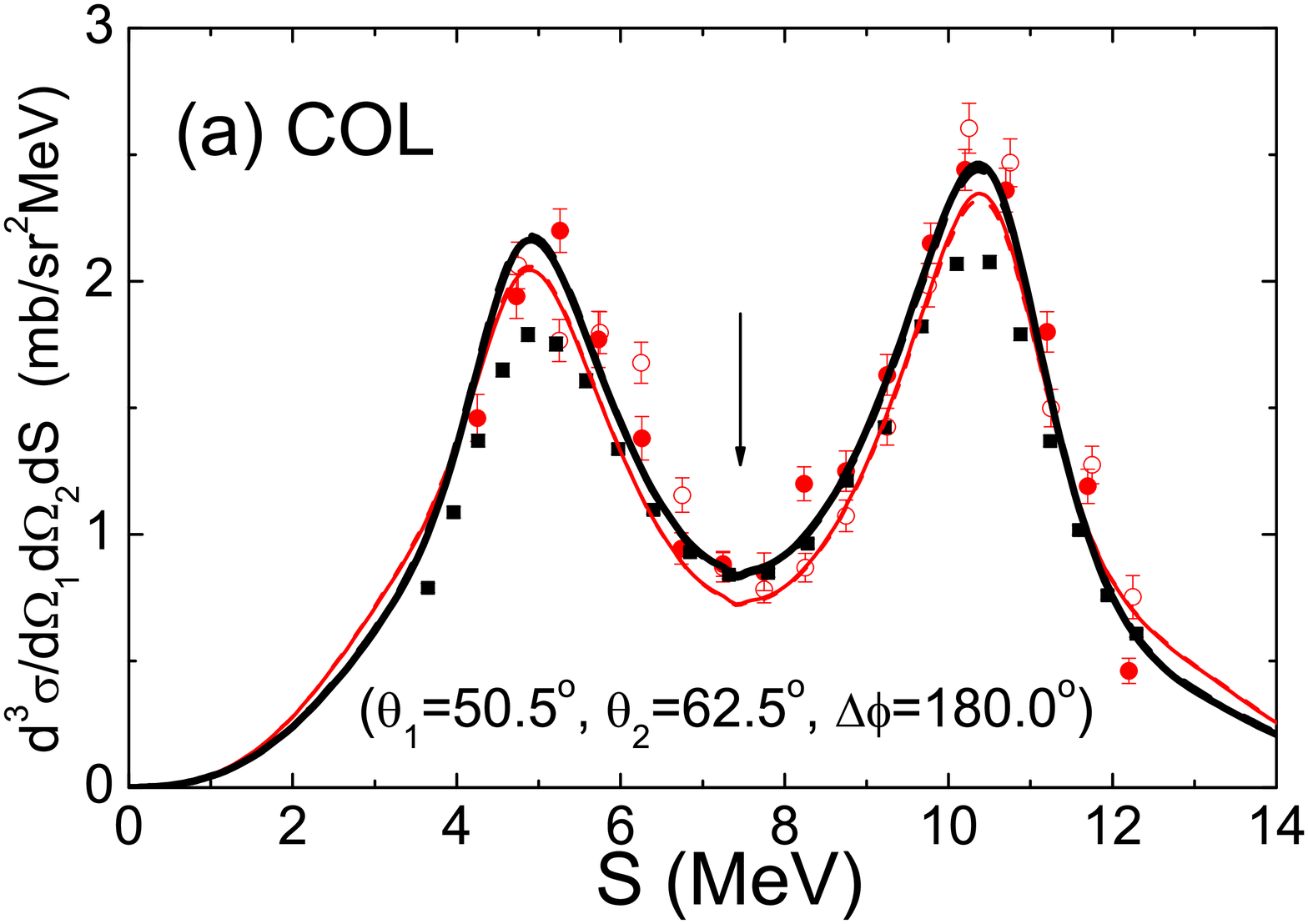}  %0.45 or 0.80
\includegraphics[width=0.80\columnwidth,angle=0]{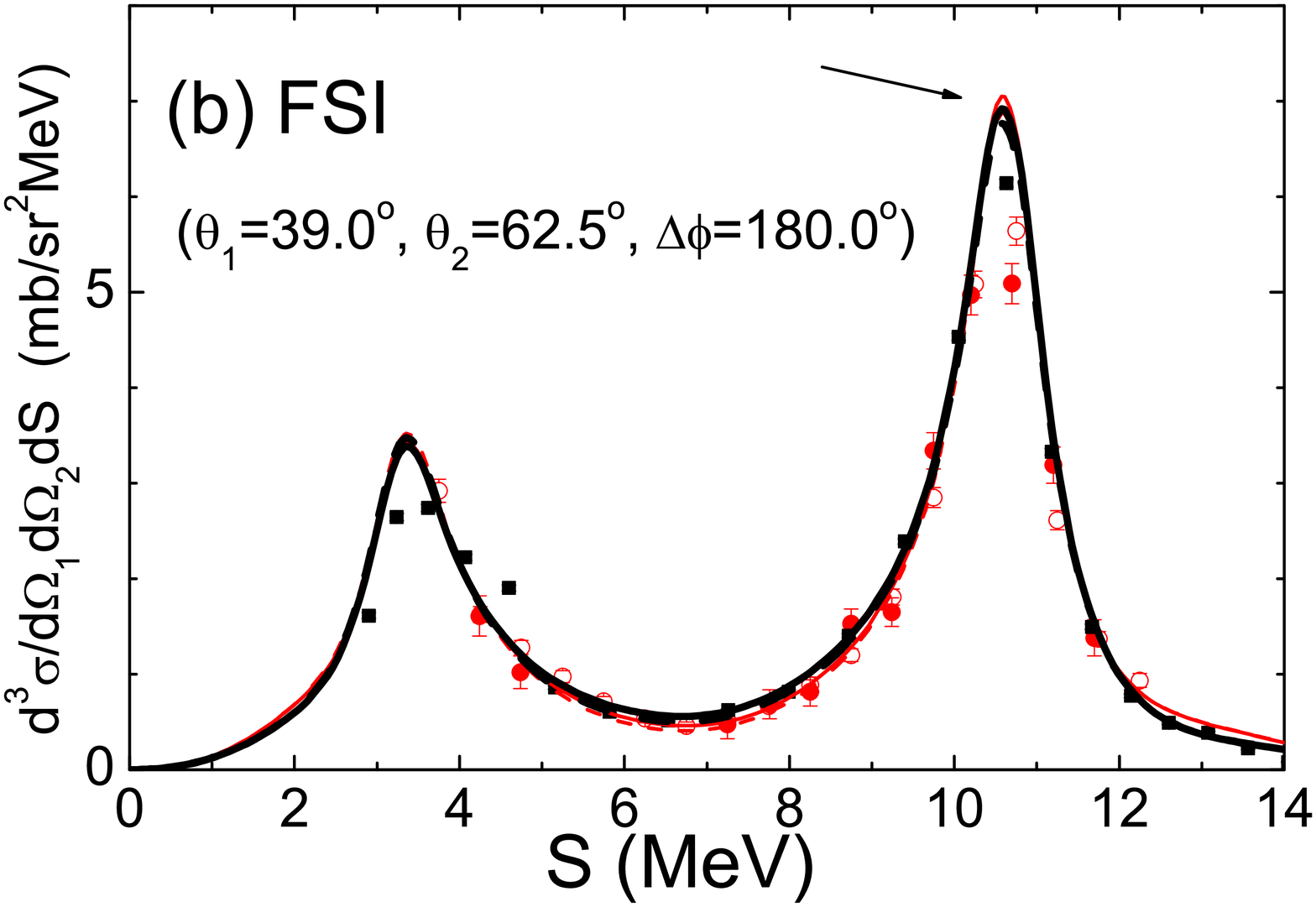}
\includegraphics[width=0.80\columnwidth,angle=0]{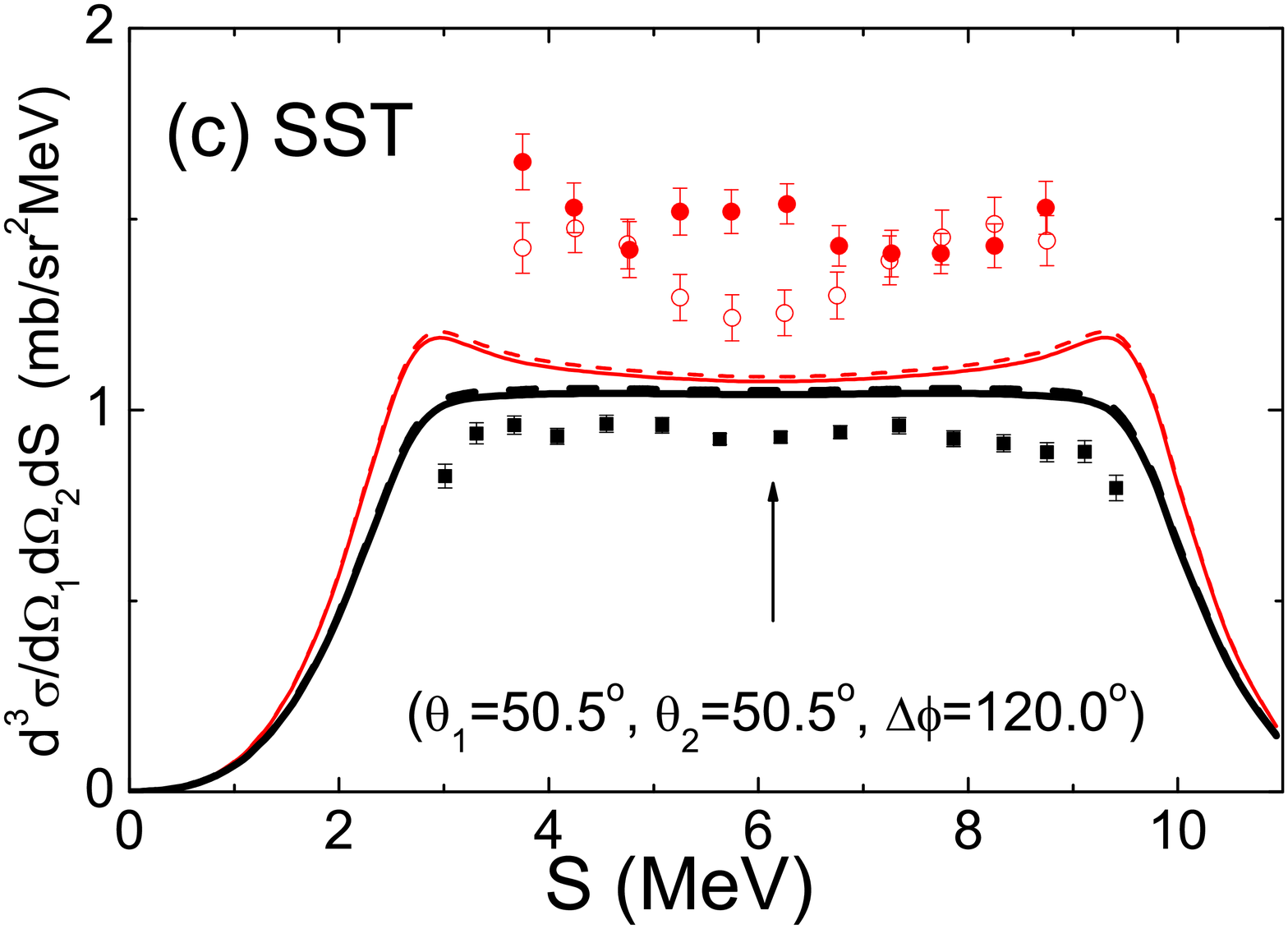}
\includegraphics[width=0.80\columnwidth,angle=0]{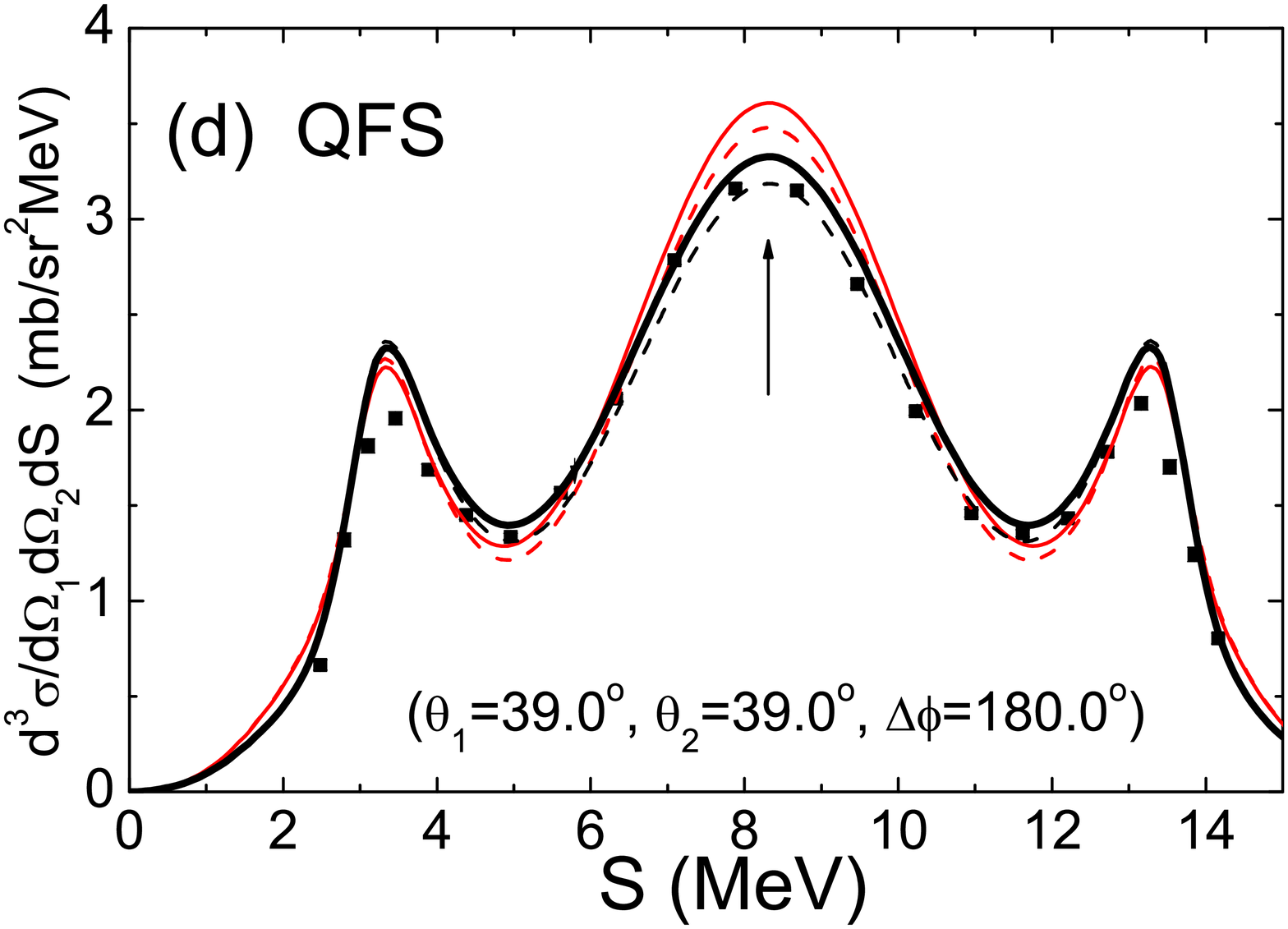}
\caption{(Color online)
Differential cross sections of $pd$ and $nd$ breakup reactions for 
(a) the COL configuration, 
(b) the FSI configuration, 
(c) the SST configuration, and (d) the QFS configuration at $E_N$ = 13.0 MeV.
The bold curves are for $pd$ and thin curves for $nd$ scattering.
The dashed curves denote the calculation with the AV18 potential, and the solid curves those with the AV18+BR$_{660}$ potential. 
Experimental data are from Ref. \cite{Ra91} (solid squares) for $pd$
and Ref. \cite{St89} (open circles) and Ref. \cite{Se05} (solid circles) for $nd$ scattering.
The arrows indicate the kinematical points that match the typical configurations.
\label{fig:breakup-13mev}}
\end{figure*}

In Fig. \ref{fig:breakup-13mev}, results of the $pd$- and $nd$-breakup cross sections for the above four configurations with the AV18 and the AV18+BR$_{660}$ potentials are compared with the experimental data. 
Effects of the Coulomb force are visible for the COL, SST, and QFS configurations, but not so for the FSI configuration, which is consistent with the result of the momentum space calculations \cite{De05b}. 
On the other hand, effects of the 2$\pi$E-3NF are small except for the QFS configurations.
In the momentum space approach \cite{De05b}, three-nucleon force effects are incorporated  alternatively in terms of an explicit introduction of a single virtual $\Delta$-isobar excitation. 
Their results also show that effects of the $\Delta$-isobar in the breakup cross sections are small for the COL, FSI, and SST configurations, and are visible for the QFS configuration.

%%%%%%%%%%%%%%%%%%%%%%%%%%%%%%%%%%%%%%%%%%%%%%%%%%%%%%%%%%%%%%%%%%%%
%%%%%%%%%%%%%%%%%%%%%%%%%%%%%%%%%%%%%%%%%%%%%%%%%%%%%%%%%%%%%%%%%%%%%%%

%%%%%%%%%%%%%%%%%%%%%%%%%%%%%%%%%%%%%%%%%%%%%%%%%%%%%%%%%%%%%%%%%%%%
%%%%%%%%%%%%%%%%%%%%%%%%%%%%%%%%%%%%%%%%%%%%%%%%%%%%%%%%%%%%%%%%%%%%%%%

%%%%%%%%%%%%%%%%%%%%%%%%%%%%%%%%%%%%%%%%%%%%%%%%%%%%%%%%%%%%%%%%%%%%
%%%%%%%%%%%%%%%%%%%%%%%%%%%%%%%%%%%%%%%%%%%%%%%%%%%%%%%%%%%%%%%%%%%%%%%

%%%%%%%%%%%%%%
%%% Sec. 5 %%%
%%%%%%%%%%%%%%
%%%%%%%%%%%%%%%%%%%%%%%%%%%%%%%%%%%%%%%%
\section{\label{sec:summary} Summary}
%%%%%%%%%%%%%%%%%%%%%%%%%%%%%%%%%%%%%%%%

We presented a practical method to solve the $pd$ scattering problem at energies above the threshold of the deuteron breakup in accommodating effects of the long-range $pp$ Coulomb force as accurate as possible.  
Although the convergence with respect to the cutoff procedure of the long-range Coulomb force effect is left as a future problem at a particular kinematical condition of breakup reactions, a successful convergence is obtained for elastic observables and for the most of kinematical regions in breakup reactions.  
We thereby calculated some observables in $pd$ and $nd$ reactions at energies up to 30 MeV. 
Effects of the two-pion exchange 3NF and the phenomenological 3NF to reproduce low-energy $3N$ observables are examined for $pd$ observables at higher energies, 
and then some discrepancies between calculations and experimental data as well as  
inconsistencies between calculations and data with respect to Coulomb force effects
are observed. 
Studies for searching realistic mechanisms to produce interaction models to remedy these defects  with calculations by the formalism presented in this paper 
are now in progress.

%%%%%%%%%%%%%%%%%%%
\begin{acknowledgements}
The author thanks H. Paetz gen Schieck for providing experimental data. 
Numerical calculations in this article are supported by Research Center for Computing and Multimedia Studies, Hosei University.
\end{acknowledgements}

%%%%%%%%%%%%%%%%
%%% Appendix %%%
%%%%%%%%%%%%%%%%
\appendix

%%%%%%%%%%%%%%%%%%%%%%%%%%%%%%%
\section{\label{sec:MCF} Method of Continued Fraction}
%%%%%%%%%%%%%%%%%%%%%%%%%%%%%%%

In this appendix, we summarize the MCF algorithm to solve the SSF equation (see Refs. \cite{Sa86,Is87} and references therein).
Let us consider to solve a linear integral equation
\begin{equation}
\left\vert \Phi \right) = 
\left\vert F \right) + {\cal G} \Delta \left\vert \Phi \right).
\end{equation}

In the notation of the present work, $\left\vert \Phi \right)$ and $\left\vert F \right)$ are expressed as vectors,
\begin{equation}
\left\vert \Phi \right) = \left( 
\begin{array}{l}
\Phi(\bm{x}_1,\bm{y}_1)
\\
\Phi(\bm{x}_2,\bm{y}_2)
\\
\Phi(\bm{x}_3,\bm{y}_3)
\end{array}
\right),
\end{equation}
\begin{equation}
\left\vert F \right) = \left( 
\begin{array}{l}
\Phi^d(\bm{x}_1) F^C(\bm{y}_1;\bm{p}_0,\eta_0)
\\
\Phi^d(\bm{x}_2) F^C(\bm{y}_2;\bm{p}_0,\eta_0)
\\
0
\end{array}
\right),
\end{equation}
and ${\cal G}$ and $\Delta$ are as matrices
\begin{equation}
{\cal G} = \left(
\begin{array}{ccc}
{\cal G}_1(E)  & 0 & 0 
\\
0 & {\cal G}_2(E) & 0 
\\
0 & 0 & {\cal G}_3(E)
\end{array}
\right),
\end{equation}
\begin{equation}
\Delta = \left(
\begin{array}{ccc}
0  & V_1^S & V_1^S 
\\
V_2^S & 0 & V_2^S
\\
V_3^S + V^C(x_3) & V_3^S + V^C(x_3) & 0
\\
  ~~~~~~~-u^C(y_1) &  ~~~~~~~-u^C(y_2) &  
\end{array}
\right).
\end{equation}

Setting $\left\vert F^{[0]} \right)$ and ${\cal G}^{[0]}$ as
\begin{eqnarray}
\left\vert F^{[0]} \right) &=& \left\vert F \right), 
\\
{\cal G}^{[0]} &=& {\cal G},
\end{eqnarray}
we define $\left\vert F^{[i]} \right)$ and ${\cal G}^{[i]}$ $(i=1,2,\dots)$ as follows:
\begin{eqnarray}
\left\vert F^{[i]} \right) &=& {\cal G}^{[i-1]} \Delta \left\vert F^{[i-1]} \right),
\label{eq:MCF-Fi}
\\
{\cal G}^{[i]} &=& {\cal G}^{[i-1]} 
\cr
&&  - \left\vert F^{[i]} \right) \frac{1}{\left( F^{[0]} \right\vert \Delta \left\vert F^{[i-1]} \right)} {\left( F^{[0]}\right\vert}
\cr
&=& {\cal G}^{[0]} 
\cr
&& - \sum_{j=1}^{i}\left\vert F^{[j]} \right) \frac{1}{\left( F^{[0]} \right\vert \Delta \left\vert F^{[j-1]} \right)} {\left( F^{[0]}\right\vert}.
\label{eq:MCF-Gi}
\end{eqnarray}

Introducing $\left\vert \Phi^{[i]} \right)$ $(i=0,1,2,\dots)$ as solutions of 
\begin{equation}
\left\vert \Phi^{[i]} \right) = \left\vert F^{[i]} \right)
 + {\cal G}^{[i]} \Delta \left\vert \Phi^{[i]} \right), 
\end{equation}
we can derive a relation between $\left\vert \Phi^{[i]} \right)$ and $\left\vert \Phi^{[i+1]} \right) $, 
\begin{equation}
\left\vert \Phi^{[i]} \right) = \left\vert F^{[i]} \right)+
\left\vert \Phi^{[i+1]} \right) \frac{\left( F^{[0]} \right\vert \Delta \left\vert F^{[i]} \right)}{\left( F^{[0]} \right\vert \Delta \left\vert F^{[i]} \right)-T^{[i+1]}}.
\label{eq:Phi-CF}
\end{equation}
Here, amplitudes $T^{[i]}$ ($i=0,1,2,\dots$) are defined as
\begin{equation}
T^{[i]} = \left( F^{[0]} \right\vert \Delta \left\vert \Phi^{[i]} \right),
\end{equation}
which satisfy
\begin{equation}
T^{[i]} = \frac{\left( F^{[0]} \right\vert \Delta \left\vert F^{[i]} \right)^2}{\left( F^{[0]} \right\vert \Delta \left\vert F^{[i]} \right)-T^{[i+1]}}.
\label{eq:T-CF}
\end{equation}

Calculations of a $N$-th order approximation start by regarding $\left\vert F^{[N]} \right)$ as $\left\vert \Phi^{[N]} \right)$: 
\begin{equation}
\left\vert \Phi^{[N]} \right) = \left\vert F^{[N]} \right) 
\end{equation}
and thereby
\begin{equation}
T^{[N]} = \left( F^{[0]} \right\vert \Delta \left\vert F^{[N]} \right).
\end{equation}
Then, using Eqs. (\ref{eq:Phi-CF}) and (\ref{eq:T-CF}) backward, we calculate 
$\left\vert \Phi^{[N-1]} \right)$, 
$\left\vert \Phi^{[N-2]} \right)$, 
$\dots$,  successively until $\left\vert \Phi^{[0]} \right)$  as the $N$-th order approximation for $\left\vert\Phi \right)$.

%%%%%%%%%%%%%%%%%%%%%%%%%%%%%%%
\section{\label{sec:functions} Coulomb Functions}
%%%%%%%%%%%%%%%%%%%%%%%%%%%%%%%

In this appendix, we summarize formulae of functions related to spectator functions modified by the Coulomb potential.
See Ref.\ \cite{Ab65} for details.

Let $F_{\ell}(\eta(p),py)$ and $G_{\ell}(\eta(p),py)$ be the regular and irregular Coulomb functions that satisfy 
\begin{equation}
\left[ T_{\ell}(y) + \frac{e^2}{y} \right] y_{\ell}(\eta(p),py) 
   = \left(\frac{3\hbar^2}{4m} p^2\right) y_{\ell}(\eta(p),py),
\label{eq:Coulomb-partial}
\end{equation}
where $y_{\ell}(\eta(p),py)$ is ether $F_{\ell}(\eta(p),py)$ or $G_{\ell}(\eta(p),py)$, and 
\begin{equation}
\eta(p) = \frac{2m}{3\hbar^2} \frac{e^2}{p}.
\label{eq:eta_p}
\end{equation}

A scattering state for the Coulomb potential $\frac{e^2}{y}$ with energy $\frac{3\hbar^2}{4m} p^2$ is written as
\begin{eqnarray}
F^c(\bm{y};\bm{p},\eta(p)) 
&=& 4\pi \sum_{\ell,m} \imath^\ell Y_\ell^{m *}(\hat{\bm{p}}) 
 Y_\ell^m(\hat{\bm{y}}) 
\cr
&& \times e^{\imath\sigma_\ell(\eta(p))} \frac{F_\ell(\eta(p),py)}{py}, 
\label{eq:Fc-pwd}
\end{eqnarray}
where $\sigma_\ell(\eta)$ is the Coulomb phase shift, 
\begin{equation}
\sigma_\ell(\eta) = {\rm arg}\Gamma(\ell+1+\imath\eta).
\end{equation}

Analytical for of the Green's function, Eq. (\ref{eq:breve-G}), is given by
\begin{eqnarray}
\breve{G}_{C,\ell}(y,y^\prime;E_{p}) &=& -\frac{4m}{3\hbar^2} p 
\frac{e^{\imath\sigma_{\ell}(\eta(p))} u^{(+)}_{\ell}(\eta(p),py_{>})}{py_{>}} 
\cr
& \times&
 \frac{F_{\ell}(\eta(p),py_{<})}{py_{<}}, 
\label{eq:brG-l}
\end{eqnarray}
where  %$F_\ell(\eta,r)$ is the (real) regular Coulomb function, 
$u^{(\pm)}_\ell(\eta,r)$ is defined as 
\begin{equation}
u^{(\pm)}_\ell(\eta,r) = 
e^{\mp\imath\sigma_\ell(\eta)} 
\left( G_\ell(\eta,r) \pm \imath F_\ell(\eta,r) \right),
\end{equation}
giving the asymptotic form as
\begin{equation}
u^{(\pm)}_\ell(\eta, r) \mathop{\to}_{r \to \infty} 
 \exp\left( \pm\imath (r - \eta \ln 2r - \ell\pi/2)\right).
\label{eq:u_c_asym}
\end{equation}
%

%%%%%%%%%%%%%%%%%%%%%%%%%%%%%%%
\section{\label{sec:Green-x}Green's operator}
%%%%%%%%%%%%%%%%%%%%%%%%%%%%%%%

In this appendix, we first review two-body Green's operators, and then describe how to calculate Eq.\ (\ref{eq:theta-Gomega}) for the case of $k=3$ and  $E_q>0$. 

We define Green's operators for the outgoing $(+)$ and the incoming $(-)$ boundary conditions with a potential consisting of a short-range potential $V^S(x)$ and a long-range Coulomb potential $V^C(x)$ as
\begin{equation}
G_{L}^{(\pm)} =  \frac1{E_q \pm \imath \varepsilon - T_{L}(x) - V^S(x) -V^C(x)}, 
\label{eq:Green_pm}
\end{equation}
\begin{equation}
G_{C,L}^{(\pm)} =  \frac1{E_q \pm \imath \varepsilon - T_L(x) -V^C(x)},
\end{equation}
which satisfy resolvent relations
\begin{equation}
G_{L}^{(\pm)} 
  = G_{C,L}^{(\pm)} + G_{L}^{(\pm)} V^S G_{C,L}^{(\pm)} 
  = G_{C,L}^{(\pm)} + G_{C,L}^{(\pm)} V^S G_{L}^{(\pm)}.
\label{eq:resolvent}
\end{equation}

Two-body scattering wave functions corresponding to the outgoing and the incoming boundary conditions $\vert \psi_L^{(\pm)}\rangle$ satisfy the (partial-wave) Lippmann-Schwinger equations
\begin{equation}
\vert \psi_L^{(\pm)}\rangle = \vert \hat{F}_L \rangle
   + G_{C,L}^{(\pm)} V^S \vert \psi_L^{(\pm)}\rangle,
\label{eq:LS-c}
\end{equation}
where $\hat{F}_L(\gamma(q),qx)$ is a reduced Coulomb function defined by
\begin{equation}
\hat{F}_L(\gamma(q),qx) \equiv \frac{F_L(\gamma(q),qx)}{qx}
\label{eq:hat-F}
\end{equation}
with
\begin{equation}
\gamma(q) = \frac{me^2}{2\hbar^2 q}.
\label{eq:gamma}
\end{equation}
For later use, we define reduced Coulomb functions $\hat{G}_L(\gamma(q),qx)$ and $\hat{u}_L^{(\pm)}(\gamma(q),qx)$ similar to Eq. (\ref{eq:hat-F}). 

Using Eq. (\ref{eq:resolvent}), we see that a formal solution of Eq. (\ref{eq:LS-c}) is written as
\begin{equation}
\vert \psi_L^{(\pm)}\rangle = \vert \hat{F}_L \rangle
   + G_{L}^{(\pm)} V^S \vert \hat{F}_L \rangle.
\label{eq:formal_sol}
\end{equation}

It is convenient to use the principal values of the two-body Green's operators defined as
\begin{equation}
{\cal P}G_{L}   = {\cal P} \frac1{E_q - T_L(x) - V^S(x) - V^C(x)},
\end{equation}
\begin{equation}
{\cal P}G_{C,L}   = {\cal P} \frac1{E_q - T_L(x)- V^C(x)}.
\end{equation}
As is $G_{C,L}^{(\pm)}$, the analytical form of ${\cal P}G_{C,L}^{(\pm)}$ is known and these operators are related as
\begin{equation}
G_{C,L}^{(\pm)} =  {\cal P}G_{C,L} \mp \imath q \frac{m}{\hbar^2} \vert \hat{F}_L \rangle \langle \hat{F}_L \vert.
\label{eq:G0PG0}
\end{equation}

A scattering wave function corresponding to ${\cal P}G_{C,L}$, namely standing wave solution $\vert\bar{\psi}_L \rangle$ satisfies
\begin{equation}
\vert \bar{\psi}_{L} \rangle = \vert \hat{F}_L \rangle 
 + {\cal P}G_{C,L} V^S \vert \bar{\psi}_{L} \rangle,
\label{eq:LS-standing}
\end{equation}
and a formal solution of this is given as 
\begin{equation}
\vert \bar{\psi}_L \rangle 
  = \vert \hat{F}_L \rangle  + {\cal P}G_{L} V^S \vert \hat{F}_L \rangle.
\label{eq:formal_sol_0}
\end{equation}

From the standing wave solution, the outgoing and the incoming solutions are obtained as
\begin{equation}
\vert \psi^{(\pm)}_{L} \rangle = 
  \frac1{1 \mp \imath {\cal K}_L } \vert \bar{\psi}_{L} \rangle,
\end{equation}
where ${\cal K}_L$ is the scattering $K$-matrix defined by
\begin{equation}
{\cal K}_{L} 
  = - q \frac{m}{\hbar^2} \langle \hat{F}_L \vert V \vert \bar{\psi}_{L} \rangle,
\end{equation}
which becomes $\tan\delta$ with a phase shift parameter $\delta$.
Using the relations above, one obtains a relation between $G_{L}^{(\pm)}$ and ${\cal P}G_{L} $ as
\begin{equation}
G_{L}^{(\pm)}  = {\cal P}G_{L} 
 \mp \imath q \frac{m}{\hbar^2} \vert \bar{\psi}_{L} \rangle 
   \frac1{1 \mp \imath {\cal K}_{L}} \langle \bar{\psi}_{L} \vert,
\label{eq:GcalG}
\end{equation}
which reduces to Eq.\ (\ref{eq:G0PG0}) if $V(x)$ was 0, leading to $\bar{\psi}_{L}(x) = \hat{F}_L(\gamma(q),qx)$ and ${\cal K}_{L} = 0$.

%%%%%%%%%%%%%%%%%%%%%%%%%%%%%%%%%%%%
Next, we discuss about asymptotic form of the Green's functions.

Using the resolvent equation (\ref{eq:resolvent}), the formal solutions, Eqs. (\ref{eq:formal_sol}) and (\ref{eq:formal_sol_0}), and the asymptotic forms of the Coulomb Green's functions, %$G_{C,L}^{(\pm)}$ and ${\cal P}G_{C,L}$, 
which are obtained from their analytical forms, 
\begin{equation}
G_{C,L}^{(\pm)} 
  \to   - q \frac{m}{\hbar^2} e^{\pm\imath\sigma_L} \vert \hat{u}_L^{(\pm)}\rangle \langle \hat{F}_L \vert,
\label{eq:G0_asym}
\end{equation}
\begin{equation}
{\cal P}G_{C,L} \to  q \frac{m}{\hbar^2} \vert \hat{G}_L \rangle \langle \hat{F}_L \vert,
\label{eq:PG0asym}
\end{equation}
we obtain the asymptotic form of the Green's functions,
\begin{equation}
G_{L}^{(\pm)} \to 
 - q \frac{m}{\hbar^2} e^{\pm\imath\sigma_L} \vert \hat{u}_L^{(\pm)}\rangle \langle \psi_L^{(\mp)} \vert, 
\end{equation}
\begin{equation}
{\cal P}G_{L}  \to  
  q \frac{m}{\hbar^2} \vert \hat{G}_L \rangle \langle\bar{\psi}_{L} \vert.
\label{eq:calG_asym}
\end{equation}

Finally, we describe how to calculate Eq.\ (\ref{eq:theta-Gomega}) for $k=3$, which we write simply as
\begin{equation}
 \theta(x) = \langle x \vert G_{L}^{(+)} \vert \omega \rangle.
\label{eq:etaGomega}
\end{equation}

Using Eq.\ (\ref{eq:GcalG}), one can write $\theta(x)$ as
\begin{equation}
\theta(x) = \bar{\theta}(x) 
- \imath q \frac{m}{\hbar^2} \bar{\psi}_{L}(x)
   \frac1{1 - \imath {\cal K}_{L}} \langle \bar{\psi}_{L} \vert \omega \rangle,
\end{equation}
where a new function $\bar{\theta}(x)$ is defined by
\begin{equation}
 \bar{\theta}(x)  = \langle x \vert {\cal P}G_{L} \vert \omega \rangle.
\end{equation}
From Eq.\ (\ref{eq:calG_asym}), the asymptotic form of $\bar{\theta}(x)$ can be written as
\begin{equation}
\bar{\theta}(x)  \mathop{\to}_{x \to \infty}  
  q \frac{m}{\hbar^2} \hat{G}_L (\gamma(q),qx) \langle\bar{\psi}_{L} \vert \omega \rangle
\end{equation}

In actual calculation, the function $\bar{\theta}(x)$ is obtained by solving the ordinary differential equation
\begin{equation}
\left[ E_q  - T_L(x) - V^S(x) - V^C(x) \right] \bar{\theta}(x)  = \omega(x)
\end{equation}
with the boundary condition
\begin{equation}
\bar{\theta}(x)  \mathop{\propto}_{x \to \infty}  \hat{G}_L (\gamma(q),qx).
\label{eq:bd-cond-3P}
\end{equation}

These relations give the asymptotic form of $\theta(x)$ as
\begin{equation}
\theta(x)  \mathop{\to}_{x \to \infty} 
 e^{+\imath\sigma_L}  \hat{u}_L^{(+)}(\gamma(q),qx) \frac1{1- \imath {\cal K}_{L}} \left( -q \frac{m}{\hbar^2} \right)
 \langle \bar{\psi}_{L} \vert \omega \rangle.
\label{eq:theta_asym}
\end{equation}

%%%%%%%%%%%%%%%%%%
%%% References %%%
%%%%%%%%%%%%%%%%%%

\end{document}